\documentclass[
aps,
prx,
showpacs,
preprintnumbers,
twocolumn,
superscriptaddress,
10pt,
longbibliography,
]{revtex4-2}

\usepackage[T1]{fontenc}
\usepackage{newtxtext}
\usepackage{newtxmath}

\usepackage{enumitem}

\usepackage{makecell}

\usepackage{physics}
\usepackage{amsmath}
\usepackage{mathtools}
\usepackage{dsfont}

\usepackage{appendix}

\usepackage{placeins}

\usepackage{multirow}
\usepackage{tabularray}

\usepackage{color}
\usepackage[dvipsnames]{xcolor}
\usepackage{colortbl}
\usepackage{hyperref}
\hypersetup{colorlinks,citecolor=blue,linkcolor=blue,urlcolor=blue}

\definecolor{darkgreen}{rgb}{0.55, 0.71, 0.00}
\definecolor{Gray}{gray}{0.9}

\usepackage{comment}

\usepackage[normalem]{ulem}

\newcommand{\specialcell}[2][c]{%
\begin{tabular}[#1]{@{}c@{}}#2\end{tabular}}

\newcommand{\K}{\mathrm{K}}
\newcommand{\F}{\mathrm{F}}

\renewcommand{\i}{\mathrm{i}}

\newcommand{\HF}{H_\F}
\newcommand{\UF}{U_\F}
\newcommand{\EF}{\varepsilon_\F}
\newcommand{\EFn}[1]{\varepsilon_{\F,#1}}

\newcommand{\EK}{\varepsilon_\K}
\newcommand{\EKn}[1]{\varepsilon_{\K,#1}}

\newcommand{\EKavgn}[1]{\overline{\varepsilon}_{\K,#1}}

\newcommand{\A}{\mathcal{A}}
\newcommand{\AF}{\A_\F}
\newcommand{\AK}{\A_\K}

\newcommand{\HK}{H_\K}
\newcommand{\HKavg}{\overline{H}_\K}
\newcommand{\AKavg}{\overline{\A}_\K}

\newcommand{\HCD}{H_\mathrm{CD}}
\newcommand{\HFCD}{H_\mathrm{F, CD}}
\newcommand{\Hctrl}{H_\mathrm{ctrl}}
\newcommand{\ctrl}{\mathrm{ctrl}}

\newcommand{\Havg}{\text{\AE}}
\newcommand{\Eavg}{\text{\ae}}

\newcommand{\identity}{\mathds{1}}

\usepackage{graphicx}
\usepackage[all]{hypcap}
\graphicspath{{visual_elements/figs/}}

\begin{document}

\clearpage
\title{Geometric Floquet theory}

\author{Paul M.~Schindler}
\email{psch@pks.mpg.de}
\affiliation{Max Planck Institute for the Physics of Complex Systems, N\"{o}thnitzer Str.~38, 01187 Dresden, Germany}
\author{Marin Bukov}
\affiliation{Max Planck Institute for the Physics of Complex Systems, N\"{o}thnitzer Str.~38, 01187 Dresden, Germany}

\begin{abstract}
    We derive Floquet theory from quantum geometry.
    We identify quasienergy folding as a consequence of a broken gauge group of the adiabatic gauge potential $U(1){\mapsto}\mathbb{Z}$. Fixing instead the gauge freedom using the parallel-transport gauge uniquely decomposes Floquet dynamics into a purely geometric and a purely dynamical evolution. The dynamical average-energy operator provides an unambiguous sorting of the quasienergy spectrum, identifying a Floquet ground state and suggesting a way to define the filling of Floquet-Bloch bands.
    We exemplify the features of geometric Floquet theory using an exactly solvable XY model and a non-integrable kicked Ising chain. We elucidate the geometric origin of inherently nonequilibrium effects, like the $\pi$-quasienergy splitting in discrete time crystals or $\pi$-edge modes in anomalous Floquet topological insulators. The spectrum of the average-energy operator is a susceptible indicator for both heating and spatiotemporal symmetry-breaking transitions.
    Last, we demonstrate that the periodic lab frame Hamiltonian generates transitionless counterdiabatic driving for Floquet eigenstates.
    This work directly bridges seemingly unrelated areas of nonequilibrium physics. 
\end{abstract}

\maketitle

\begin{figure*}[t]
    \centering
    \includegraphics[width=\linewidth]{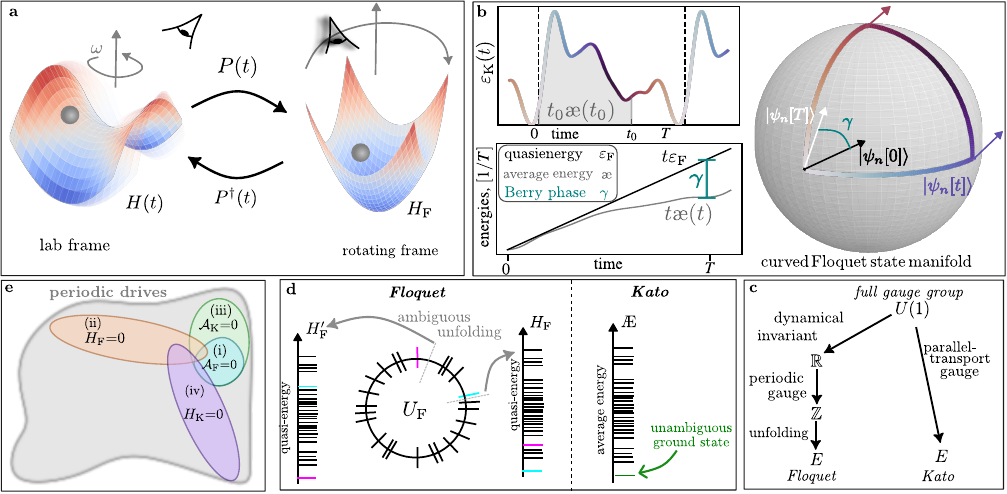}
    \caption{
        \textbf{Quantum geometric perspective on Floquet theory.}
        \textbf{a}, Floquet's theorem postulates that, for every periodically-driven system $H(t)$ (left), there exists a transformation $P(t)=P(t+T)$ to a rotating frame (right), where the dynamics are governed by a time-independent Floquet Hamiltonian~$\HF$, which can be very different from the lab-frame Hamiltonian~$H(t)$.
        \textbf{b}, the Kato decomposition of Floquet's theorem presents a geometric interpretation: as the initial time/phase of the drive is varied, Floquet states $\ket{\psi_n[t]}$ form closed trajectories on a curved state manifold. Time evolution over one drive cycle corresponds to transporting the Floquet states along the trajectory, $\ket{\psi_n[0]}{\to}\ket{\psi_n[T]}{=}\ket{\psi_n[0]}$; likewise, the phase $\theta_\F=T\EF$ accumulated after one cycle can, in turn, be decomposed into a geometric contribution, $\gamma$, and a dynamic contribution $\Eavg$.
        \textbf{c}, gauge-ambiguity and gauge-fixing in Floquet theory: changing the global phase of quantum states is a $U(1)$ gauge transformation of $\A_\F$ with no physically observable consequences. The full gauge group $U(1)$ of local time-periodic functions is broken to the group of (global) constant shifts by demanding dynamical invariance, and further broken to $\mathbb{Z}$ by imposing the periodic gauge, $\mathcal{T}\exp(-i\int_0^T \A_\F(t) \mathrm{d}t){=}\identity$; the remaining gauge freedom corresponds to the ambiguity in defining quasienergy levels $\EF{+}\ell \omega$, $\ell {\in} \mathbb{Z}$. Instead, the unique Kato decomposition uses the geometric parallel-transport gauge.
        \textbf{d}, a central challenge in Floquet theory is the ambiguous unfolding of the Floquet phases $\theta_\F$ to obtain the Floquet Hamiltonian $\HF$. By contrast, the average energies $\Eavg(T)$ present an unambiguous sorting of the Floquet states, which uniquely defines a Floquet ground state.
        \textbf{e}, distinctive families of periodic drives: (i) equilibrium drives, (ii) pure-micromotion drives, (iii) flat drives, and (iv) pure-geometric drives, help build minimal models of periodically driven systems, cf.~Discussion \& outlook.
    }
    \label{fig:introduction}
\end{figure*}

\textit{Introduction.---}%
Periodic motion is both ubiquitous and fundamental in nature, and encompasses a plethora of oscillatory phenomena ranging from the propagation of photons, and elementary particles accelerated along cyclotron orbits, to the cosmic motion of planets and galaxies. 
The general theory of periodically driven systems dates back to 19'th century mathematics and the study of small perturbations to closed orbits in phase space, culminating in the celebrated Floquet theorem~\cite{Floquet1883}. 
Physically, the theorem states that, for every periodically driven system (described by a Hamiltonian $H(t){=}H(t{+}T)$ with period $T$), there exists a \textit{rotating} reference frame~\footnote{i.e., one that coincides with the lab frame every period $T$ (stroboscopically): $P(t){=}P(t{+}T)$.}, in which the dynamics is generated by an effective time-independent Hamiltonian (so-called Floquet Hamiltonian, $\HF$), cf.~Fig.~\ref{fig:introduction}a. 
It generalizes the idea of a static observer positioned within a co-rotating frame like a Merry-go-round or the Earth.
Denoting the transformation from the lab to this rotating frame by $P(t){=}P(t{+}T)$, Floquet's theorem can be mathematically formulated as
\begin{equation}
\label{eq:Floquet}
    \HF = P^\dagger(t) H(t) P(t)- P^\dagger(t)i\partial_t P(t).
\end{equation}
The first term on the right-hand side describes the rotated system; the second describes potentials due to fictitious forces, generalizing the concept of centrifugal and Coriolis force. 

Over the last decade, Floquet theory has proven instrumental in disparate branches of physics, advancing our understanding of dynamical instabilities~\cite{Bittani2009_periodicbook,mathieu1868_memoire,Arscott1964_Mathieu,kovacic2018mathieu}, thermalization~\cite{abanin2015_heating,Moessner2017_FloquetMatter,Weidinger2017,Mori2018_prethermalization,Abanin2017_prethermalization,Takashi2016_HeatingBounds,Haldar2018_FloquetHeating}, energy bands in materials~\cite{Bloch1929}, and light-atom interactions~\cite{CohenTannoudji1998_AtomLight,Wu2007_AtomLight,Shirley1965_FloquetTheory,Sambe1973_AtomLightPeriodic}. 
Recently, periodically driven systems have seen a resurgence of interest in quantum simulation; a central aspect is the ability to engineer fictitious forces which vastly enriches the variety of phenomena that can be simulated with limited control. This so-called \textit{Floquet engineering}, i.e., the ability to implement a desired Floquet Hamiltonian $\HF$ by engineering the corresponding lab-frame drive $H(t)$, has emerged as an indispensable toolbox in quantum simulation allowing to ascribe tailored properties to quantum matter~\cite{Bukov_2015_general_HFE,Oka2019_FloquetMaterials,Cooper2019_UltracoldAtomTopoSummary,Rudner2020_FloquetTopo}.

Despite this progress, many fundamental questions remain unanswered. For instance, an ambiguity in defining a unique rotating frame strongly limits the physical interpretation of the Floquet Hamiltonian, cf.~Fig.~\ref{fig:introduction}d: quantum-simulating ground state physics or filling of energy bands at a fixed chemical potential is a priori ill-posed, yet highly desirable.
Moreover, the explicit numerical calculation of $\HF$ is a computationally challenging problem for many-body quantum systems since it requires solving exponentially many (in the number of particles) differential equations.
So far, known approximation techniques rely predominantly on perturbative expansions in the small-period regime $T{\to}0$; however, these expansions are known to diverge, restricting the analytical handle on periodically driven systems~\cite{dallesio2013_heating,dallesio2014_heating,Eckardt2015_VanVleck,dalessio2016_heating,Bukov2016_resonances}.

We address these challenges by providing a formulation of Floquet's theorem based on concepts from quantum geometry, like the adiabatic gauge potential (AGP) $\A$ and the geometric phase $\gamma$. We first prove that the lab frame Hamiltonian $H(t)$ generates transitionless driving for Floquet eigenstates, and plays the role of a counterdiabatic (CD) Hamiltonian~\cite{Berry2009_CD,Demirplak2003_CD,Demirplak2005_CD,Demirplak2008_CD,delCampo2013_CD,Jarzynski2013_STA,Kolodrubetz2017_GeometryReview,GuryOdelin2019_STA}
\begin{equation}
\label{eq:Floquet_STA}
    H(t) = \HF[t] + \A_\F(t) \equiv \HFCD(t),
\end{equation}
where the gauge potential $\A_\F(t){=}i\partial_t P(t) P^\dagger(t)$ suppresses excitations between eigenstates of $\HF[t]{=}P(t)\HF[0]P^\dagger(t)$; $[t]$ denotes the initial time/phase of the drive (a.k.a.~Floquet gauge). 
Equation~\eqref{eq:Floquet_STA} implies that finding the Floquet Hamiltonian is an inverse counterdiabatic driving problem, providing means to \textit{explicitly} find the Floquet Hamiltonian $\HF[t]$ for a given lab-frame Hamiltonian $H(t)$~\cite{Sels2017_LCD,GuryOdelin2019_STA,Claeys2019_FloquetCDProtocols,takahashi2023_Krylov}.
This perspective allows us to derive Floquet theory from the adiabatic theorem.

Further, we identify quasienergy folding as a consequence of a broken gauge group $U(1){\mapsto}\mathbb{Z}$ of the gauge potential $\A_\F$, with the remaining gauge freedom corresponding to the ambiguity in defining quasienergy levels $\EF{\mapsto}\EF {+}m \omega$, $m {\in} \mathbb{Z}$ ($\omega{=}2\pi/T$).
We then introduce an alternative formulation of Floquet theory based on the parallel-transport gauge that decomposes the evolution into a geometric part described by a Wilson line operator, and a dynamical part governed by an average-energy operator whose eigenstates are the Floquet states. This enables us to unambiguously sort the quasienergy spectrum, and uniquely identify the \textit{Floquet ground state} at any drive frequency. 

For the kicked Ising chain, we find a critical frequency at which the average-energy operator changes from a $2$-local to a nonlocal operator; as a result, the corresponding many-body bandwidth loses extensivity and exhibits `spectral implosion', offering new insights into the breakout of heating. 
Remarkably, we reveal quantum geometry as the driver of inherently nonequilibrium effects: we trace back the origins of anomalous Floquet topological insulators--in agreement with previous studies~\cite{gavensky2024stredaformulafloquetsystems}--and discrete time crystals to the geometric phase.
Geometric phases exhibit strong susceptibility to the presence of nonequilibrium phase transitions and rapid changes when crossing Floquet resonances, offering a sensitive probe for the critical parameter strength. 
Therefore, Floquet geometry emerges as a new tool to identify and classify nonequilibrium phenomena.

\begin{table}[t]
    \centering
    \begin{tabular}{|c||c|c|}
    \hline
        \textit{drive}
       & \textit{Hamiltonian} $H(t)$ 
       & \textit{accumulated phase} 
       \\
       \hline
       \hline
       \specialcell{adiabatic\\ $T_\mathrm{ramp}{\to}\infty$ }
       & $\Hctrl$
       & $\gamma_n(t) + \phi_n(t)$
       \\
       \hline
       \specialcell{Kato counterdiabatic \\ parallel-transport gauge}
       & $\Hctrl + \A_{\K,\lambda}$
       & $\gamma_n(t) + \phi_n(t)$
       \\
       \hline
       \specialcell{generic counterdiabatic \\ $\chi_n$ arbitrary}
       & $\Hctrl + \A^\prime_\lambda$
       & $\chi_n(t) {+} \gamma_n(t){+}\phi_n(t)$
       \\
       \hline
       \specialcell{dyn.~counterdiabatic \\ $\chi_n(t){=}{-}\gamma_n(t)$}
       & $\Hctrl + \A_{\mathrm{D},\lambda}$
       & $\phi_n(t)$
       \\
       \hline
       \specialcell{Kato AGP \\ parallel-transport gauge}
       & $\A_{\K,\lambda}$
       & $\gamma_n(t)$
       \\
       \hline
       \hline
       \specialcell{periodic AGP \\ $\chi_n(t){=}{-}\gamma_n(t){+}2\pi\ell_n t/T$ 
       }
       & $\A_{\F}(t){=}\A_{\F}(t{+}T)$
       & $2\pi\ell_n$, $\ell_n{\in}\mathbb{Z}$ (at $t{=}T)$\\
    \hline
    \end{tabular}
    \caption{\textbf{Adiabatic gauges}. 
    Summary of common gauge choices and the resulting accumulated phases for the Schr\"odinger equation $i\partial_t\ket{\psi_n(t)}{=}H(\lambda(t))\ket{\psi_n(t)}$.
    Irrespective of the gauge choice for the \textit{adiabatic gauge potential}~(AGP) $\A$, the AGP induces transitionless driving between eigenstates of $\Hctrl$; however, the gauge choice determines the accumulated phase [see text]. 
    The dynamical phase is $\phi_n(t)$, the geometric phase is $\gamma_n(t)$, and $\chi_n(\lambda(t))$ is an arbitrary smooth function; the periodic gauge is only well-defined for periodic control with $\ell_n{\in}\mathbb{Z}$.
    For proof of these relations, see appendix~\ref{app:proof_table}.
    }
    \label{tab:gauges}
\end{table}


\textit{Adiabatic evolution revisited.---}%
To understand the meaning of Eq.~\eqref{eq:Floquet_STA}, let us start by providing a brief overview of adiabatic control and shortcuts to adiabaticity~(STA)~\cite{Berry2009_CD,Demirplak2003_CD,Demirplak2005_CD,Demirplak2008_CD,delCampo2013_CD,Jarzynski2013_STA,Kolodrubetz2017_GeometryReview,Sels2017_LCD,Xi2010_FewLevelSTA,campbell2015shortcut,GuryOdelin2019_STA,orozcoruiz2024_AlgebraOptimalControl}. To this end, consider a Hamiltonian $\Hctrl(\lambda)$ with some control parameter $\lambda$, and (nondegenerate) eigenstates $\Hctrl(\lambda)\ket{\psi_n[\lambda]}=E_n(\lambda)\ket{\psi_n[\lambda]}$. 
If $\lambda(t)$~($0{\leq} t {\leq} T_\mathrm{ramp}$) is varied infinitely slowly in time, $\dot{\lambda}{\to} 0$ with $T_\mathrm{ramp}{\to}\infty$, the adiabatic theorem~\cite{Born1928_adiabatic,Kato1950_adiabatic} guarantees that the evolution follows the instantaneous eigenstates, $\ket{\psi_n[\lambda]}$, of the control Hamiltonian, $\Hctrl(\lambda)$:
\begin{eqnarray}
    \label{eq:adiabatic}
    \ket{\psi(t)} &=& 
    \mathcal{T}\exp\left(-i\int^t_0 \Hctrl(\lambda(s))\mathrm ds \right) \ket{\psi_n[\lambda(0)]} \nonumber\\
    &\to&
    e^{i\gamma_n(t)} e^{i\phi_n(t)} \ket{\psi_n[\lambda(t)]},
\end{eqnarray}
with the dynamical phase $\phi_n(t){=}-\int_0^t E_n(\lambda(s)) \dd s$, and the geometric phase $\gamma_n(t)$=$-\int_{\lambda(0)}^{\lambda(t)} \braket{\psi_n[\lambda]}{i \derivative{\lambda} \psi_n[\lambda]} \dd \lambda$ which only depends on the trajectory traced in Hilbert space by $\ket{\psi_n[\lambda]}$.

Away from the infinitely slow driving regime, diabatic transitions between eigenstates are induced during the ramp.
The goal of shortcuts to adiabaticity is to generate time-evolution that follows the adiabatic trajectory of states in Hilbert space away from the adiabatic limit~\cite{GuryOdelin2019_STA}; this requires finite modifications to the protocol $\Hctrl(\lambda){\to} \Hctrl(\lambda) + H_\mathrm{mod}$.
In particular, \textit{counterdiabatic driving}~(CD) achieves this by exactly removing all diabatic transition matrix elements~\cite{Demirplak2003_CD,Demirplak2005_CD,Demirplak2008_CD,Berry2009_CD}. To identify these transitions, we consider a transformation $V_\lambda$ to the instantaneous eigenbasis  of $\Hctrl(\lambda)$; in the corresponding co-moving frame, the dynamics are governed by
$
    \Tilde{H}_\mathrm{ctrl}(\lambda) {-} \dot{\lambda} \Tilde{\A}_\lambda
$
with the \textit{adiabatic gauge potential}~(AGP) $\A_\lambda {=} i \pqty{ \partial_\lambda V_\lambda}V^\dagger_\lambda$, and $\Tilde{(\cdot)}{=}V_\lambda^\dagger (\cdot) V_\lambda$. By construction, $\Tilde{H}_\mathrm{ctrl}(\lambda)$ is diagonal, such that all excitations between its eigenstates in the co-moving frame are described by the AGP $\Tilde{\A}_\lambda$. Therefore, transitionless driving is obtained by removing the transition matrix elements in the lab frame:
$
    H_\mathrm{CD}(\lambda) = \Hctrl(\lambda) + \dot{\lambda} \A_\lambda \, .
$
The AGP $\A_\lambda$ itself does not contain the information to diagonalize the Hamiltonian $\Hctrl(\lambda)$, but rather describes the infinitesimal transformation between instantaneous eigenbases~\footnote{The information about the eigenstates is contained in $V(\lambda(0))$ -- the initial condition to the Schroedinger equation $i\partial_\lambda V(\lambda) = \A_\lambda V(\lambda)$}; this, is evident from its matrix elements 
$
    \bra{\psi_m}\A_\lambda\ket{\psi_n} = i\braket{\psi_m}{\partial_\lambda \psi_n}.
$

The AGP $\A_\lambda$ is not uniquely defined due to the $U(1)$ gauge transformation
\begin{eqnarray}
    \A_\lambda &\mapsto& \A_\lambda^\prime = \A_\lambda {-} \sum_n \partial_\lambda\chi_n(\lambda) \ketbra{\psi_n} \notag\\
    \ket{\psi_n[\lambda]} &\mapsto& \ket{\psi^\prime_n[\lambda]}=e^{i\chi_n(\lambda)} \ket{\psi_n[\lambda]},
\end{eqnarray}
which corresponds to choosing the overall phase of each individual eigenstate: $\ket{\psi^\prime_n[\lambda]}$ is also a valid eigenbasis of $\Hctrl(\lambda)$ for a fixed $\lambda$.
Indeed, we can add to the AGP arbitrary operators, diagonal in the eigenbasis $\ket{\psi_n[\lambda]}$, to obtain another valid AGP; while all AGPs lead to transitionless driving of the eigenstates, the accumulated phases during the evolution may differ (see Table~\ref{tab:gauges}). As a result, evolution of eigenstates $\ket{\psi_n[\lambda(0)]}$ under $H^\prime_\mathrm{CD}(\lambda) {=} \Hctrl(\lambda) {+} \dot{\lambda} \A^\prime_\lambda$ obeys
$
    \label{eq:agp_evolution}
    \ket{\psi(t)} {=} e^{i\chi_n(\lambda(t))} e^{i\gamma_n(t)} e^{i\phi_n(t)} \ket{\psi_n[\lambda(t)]}.
$
However, note that non-eigenstates undergo different evolution under $H_\mathrm{CD}$ and $H^\prime_\mathrm{CD}$.

We can remove the gauge ambiguity by considering the adiabatic \textit{parallel-transport} gauge defining the \textit{Kato} AGP~\cite{Kato1950_adiabatic,budich2013adiabatic}
\begin{equation}
    \label{eq:kato_agp}
    \A_{\mathrm{K},\lambda}=-\frac{1}{2}\sum_n \comm{\Pi_n[\lambda]}{i\partial_\lambda \Pi_n[\lambda]},
\end{equation}
expressed in terms of the commutator of the gauge-invariant eigenstate projectors $\Pi_n(\lambda){=}\ketbra{\psi_n[\lambda]}$.
The corresponding Kato CD driving is the \textit{unique} gauge choice reproducing the adiabatic evolution~\eqref{eq:adiabatic} exactly -- including the correct phase evolution of eigenstates -- lifting the constraint of the adiabatic regime~\footnote{However, the condition on the spectrum $E_n(\lambda)$ of $\Hctrl(\lambda)$ being nondegenerate and gapped persists.}, cf.~Table~\ref{tab:gauges}.
One can show that its matrix elements can be obtained from those of $\A_\lambda$ by removing all diagonal entries, $\bra{\psi_m}\A_{\K,\lambda}\ket{\psi_n} {=} (1 {-} \delta_{nm})\braket{\psi_m}{i\partial_\lambda \psi_n}$.
Analogous to the dynamical-gauge AGP $\A_\lambda$ being interpreted as the derivative operator $\A_\lambda\hat{=}i\partial_\lambda$, the Kato AGP has the meaning of a covariant derivative. 
Geometrically, evolution under $\A_{\K,\lambda}$ alone implements parallel transport of the eigenstates $\ket{\psi[\lambda]}$, thus, accumulating only a geometric phase $\gamma_n$~\cite{Kato1950_adiabatic}, cf.~Table~\ref{tab:gauges}.

For generic many-body systems, the AGP is a nonlocal operator~\cite{Saberi2014_NonlocalAGP,Hatomura2018_STAnonlocalClassical,Kolodrubetz2017_GeometryReview,Sels2017_LCD}. Finding approximations without having access to spectral properties has been investigated in great depth~\cite{Saberi2014_NonlocalAGP,Sels2017_LCD,Ieva2023_COLD,morawetz2024efficient,GuryOdelin2019_STA}, including using variational~\cite{Sels2017_LCD,Ieva2023_COLD,morawetz2024efficient} and Krylov space methods~\cite{Claeys2019_FloquetCDProtocols,takahashi2023_Krylov}. One of the key contributions of this work is to pave the way to using techniques from shortcuts to adiabaticity to compute the Floquet Hamiltonian.


\textit{Geometric Floquet theory.---}%
Consider a periodic change in a parameter, $\lambda\hat{=}t$, of a control Hamiltonian $\Hctrl(t){=}\Hctrl(t{+}T)$ with non-degenerate eigenstates~(for degenerate eigenstates see App.~\ref{app:degenerate}).
We will now derive Floquet's theory using counterdiabatic driving; in Table~\ref{tab:concepts} we summarize the different concepts used in the derivation and proceeding discussion.

Due to the uniqueness of the Kato AGP, there exists a unique CD Hamiltonian $\HCD(t) {=} \Hctrl(t) {+} \A_\K(t)$, which is also periodic since both $\Hctrl(t)$ and $\A_\K(t)$ are periodic with $T$.
Crucially, the converse is also true: any periodic Hamiltonian $H(t)$ can be decomposed in the form 
\begin{equation}
\label{eq:Kato_floquet}
    H(t)=H_\K(t) + \A_\K(t) \, ,
\end{equation}
such that $H(t)$ is the counterdiabatic Hamiltonian generating transitonless driving of the eigenstates $\ket{\psi_n[t]}$ of a unique, periodic \textit{Kato Hamiltonian} $H_\K(t)$, and $\A_\K(t)$ is the corresponding periodic Kato AGP~(cf.~proof in appendix~\ref{app:proof_kato}).
Despite the similarity between the Kato decomposition Eq.~\eqref{eq:Kato_floquet} and the modified Floquet theorem Eq.~\eqref{eq:Floquet_STA}, there are two notable distinctions:
(i) $H_\K$ is not a dynamical invariant, $i\partial_t \HK(t){\neq}[\HK(t),H(t)]$, i.e., it may have time-dependent eigenvalues $\EKn{n}(t)$;
(ii) $\A_\K$ does not generate a proper rotating frame transformation, i.e., $\mathcal{T}\exp(-i \int_0^t\A_{\K,\lambda}(s)\mathrm ds)$ may not be periodic in $t$. 

To derive the familiar Floquet theorem, we change the $U(1)$-gauge for the AGP: we add and subtract a term $\mathcal{D}$ diagonal in the Kato eigenbasis $\ket{\psi_n[t]}$, leading to $\HK(t){+}\A_\K(t){=}({H}_\K(t){+}\mathcal{D}(t)){+}(\A_\K(t){-}\mathcal{D}(t))$. 
We now show that both conditions (i) and (ii) can be satisfied in two steps, $\mathcal{D}{=}\mathcal{D}^\text{(i)}{+}\mathcal{D}^\text{(ii)}$, respectively. 
First, we set $\mathcal{D}^\text{(i)}(t){=}{-}\sum_n \partial_t\chi_n(t) \ketbra{\psi_n[t]}$, where $\partial_t\chi_n(t){=}\sum_{\ell \neq 0} \EKn{n}^{(\ell)} e^{i\ell \omega t}$ and $\EKn{n}^{(\ell)}$ are the Fourier coefficients of the $T$-periodic Kato eigenenergies; by construction, this cancels all but the zeroth harmonic in the eigenenergies of $\HK^\text{(i)}[t]{=}{H}_\K(t){+}\mathcal{D}^\text{(i)}(t)$, and makes its spectrum time-independent. Therefore, $\HK^\text{(i)}[t]$ is now a dynamical invariant~\footnote{We use the notation $H_\K[t]$ to explicitly denote that the spectrum of the operator $H_\K$ is $t$-independent, while its eigenstates are not}, whereas the AGP $\AK^\text{(i)}(t){=}\AK(t){-}\mathcal{D}^\text{(i)}(t)$ has periodic diagonal elements that period-average to zero. 
Second, we add and subtract the time-independent Berry phases, $\mathcal{D}^\text{(ii)}(t){=}\sum_n T^{-1}\gamma_n(T) \ketbra{\psi_n[t]}$, leading to $\HF[t] {=} H_\K^\text{(i)}[t] {+}\mathcal{D}^\text{(ii)}(t)$ and $\A_\F(t) {=} \A_\K^\text{(i)}(t) {-} \mathcal{D}^\text{(ii)}(t)$~\footnote{Note that this second shift is not unique since the Berry phase is only defined up to shifts by an integer multiple of $2\pi$, in agreement with the quasienergy folding problem.}; by construction, $\A_\F(t)$ now generates a periodic unitary, which we identify with the micromotion operator $P(t){=}\mathcal{T}\exp(-i\int_0^t \A_\F(s) \dd s){=}P(t{+}T)$, cf.~Table~\ref{tab:gauges}. 
This completes the quantum geometric proof of the Floquet theorem~\eqref{eq:Floquet_STA}: $H(t){=}\HF[t] {+} \A_\F(t)$; together with $\HF[t] {=} P(t) \HF[0] P^\dagger(t)$, this relation implies Eq.~\eqref{eq:Floquet}.

Since $\HK(t)$ and $\HF[t]$ differ only by diagonal terms in the Kato eigenbasis, it follows that $\EFn{n}{=}T^{-1}\left(\int_0^T \EKn{n}(t)\mathrm dt{-}\gamma_n(T)\right)$, and therefore the Kato eigenstates are the Floquet states. Therefore, the lab frame Hamiltonian $H(t)$ generates transitionless (CD) driving for the Floquet eigenstates w.r.t.~changing the initial time/phase of the drive. Moreover, it is now clear that the Floquet problem is an inverse CD problem: the CD Hamiltonian $H(t)$ is given, and one looks for the corresponding control Hamiltonian $\HF[t]$. In fact, the CD formulation of Floquet's theorem, Eq.~\eqref{eq:Floquet_STA}, is naturally suited to Floquet engineering applications, where one starts with a desired Floquet Hamiltonian $\HF[t]$, and seeks to find the lab frame Hamiltonian $H(t)$ that implements it. 

The steps in the derivation of the CD formulation of Floquet's theorem allow us to directly identify quasienergy folding as a consequence of a broken gauge group for the AGP $U(1){\mapsto}\mathbb{Z}$. To see this, consider a generic gauge transformation with a real-valued phase $\chi_n(t)$; imposing time-periodicity of the wavefunction requires $\exp\left[i\chi_n(t{+}T)\right]{=}\exp\left[i\chi_n(t)\right]$. The most general such periodic function can be expanded as $\chi_n(t){=}m\omega t {+} \sum_{\ell\neq 0} \chi_n^{(\ell)}\exp(i\ell\omega t)$ for some $m{\in}\mathbb{Z}$. Applied to the right-hand side of Eq.~\eqref{eq:Floquet_STA}, this gauge transformation adds a diagonal term ${\propto}\partial_t\chi_n{=}m\omega{+} \sum_{\ell\neq 0} i \ell\omega\chi_n^{(\ell)}\exp(i\ell\omega t)$ to the Floquet Hamiltonian $\HF[t]$. Imposing that the quasienergies be time-independent requires that all Fourier components vanish, $\chi_n^{(\ell)}{=}0$, reducing the gauge freedom to $\chi_n(t){=}m\omega t$. Note that this amounts to an incomplete gauge fixing; the residual $\mathbb{Z}$ gauge freedom can be fixed by selecting a specific unfolding gauge, cf.~Fig.~\ref{fig:introduction}c. 

\begin{table}[t]
    \centering
    \begin{tblr}{
        colspec = {|p{0.34\linewidth}|p{0.62\linewidth}|},
        vlines,
    }
    \hline
        \centering \textit{Operator}
        & 
        \hspace{.36\linewidth}\textit{Definition}
        \\
        \hline
        \hline
        \centering
        Control Hamiltonian, $\Hctrl(\lambda)$
        & defines adiabatic trajectory~\eqref{eq:adiabatic} via its instantaneous eigenstates $\ket{\psi_n[\lambda]}$ 
        \\
        \hline
        \centering
        Adiabatic Gauge Potential (AGP), $\mathcal{A}_\lambda$ 
        & infinitesimal generator of instantaneous eigenbasis transformations, $\bra{\psi_m}\A_\lambda\ket{\psi_n} = i\braket{\psi_m}{\partial_\lambda \psi_n}$, with gauge-dependent diagonal elements, cf.~Table~\ref{tab:gauges}
        \\
        \hline
        \hline
        \centering
        Floquet Hamiltonian, $\HF[t]$~\eqref{eq:Floquet}
        & effective Hamiltonian, for starting time $t$, describing stroboscopic dynamics of a periodically driven system, with ambiguous quasienergies~(see Fig.~\ref{fig:introduction}d)
        \\
        \hline
        \centering
        Kato Hamiltonian, $\HK(t)$~\eqref{eq:Kato_floquet}
        & describes adiabatic evolution~\eqref{eq:adiabatic} of Floquet states, with unambiguously sorted energies and the same eigenstates as the Floquet Hamiltonian
        \\
        \hline
        \centering
        Average-energy operator, $\Havg[t]$
        & invariant of motion, with eigenvalues corresponding to the period averaged Kato energies $\Eavg_n{=}T^{-1}\int_0^T \EKn{n}(t)\dd t$ and the same eigenstates as the Floquet Hamiltonian.
        \\
    \hline
    \end{tblr}
    \caption{
        \textbf{Central definitions.}
        Summary of key operators and their corresponding definitions used throughout this manuscript; for details on the different gauges for adiabatic gauge potentials, see Table~\ref{tab:gauges}.
    }
    \label{tab:concepts}
\end{table}

By contrast, the Kato decomposition~\eqref{eq:Kato_floquet}, which ensures parallel transport, is an alternative to the Floquet decomposition~\eqref{eq:Floquet_STA}, and does not suffer from the folding ambiguity. 
Physically, the Kato Hamiltonian $\HK(t)$ is unique in that it describes the correct adiabatic evolution of Floquet eigenstates in the sense of Eq.~\eqref{eq:adiabatic}, and reproduces the correct adiabatic limit ($T{\to} \infty$, $\omega{\to} 0$ with $\omega T{=}2\pi$ fixed, and slowly changing $t$); this means that $\A_\K{\to}0$ always as $T{\to}\infty$, unlike $\A_\F$.
However, let us reiterate that $\HK(t)$ and $\HF[t]$ share the same eigenstates, which ultimately determine geometric and topological properties, such as Thouless pumping~\cite{citro2023thouless}.
We emphasize that, since the parameter we vary is the initial time/phase of the drive, no gaps can open up in the Kato spectrum, and any energy level crossings are exact; in particular, no (continuous) phase transitions can occur and hence the AGP is well-defined at all points along the path.

However, unlike the Floquet decomposition, the Kato formulation does not give rise to proper micromotion evolution; instead, we find
\begin{equation}
    \label{eq:Kato_evo}
    U(t,0) = \mathcal{T} \exp(-i\int_0^t H(s) \dd s) = e^{i\Gamma(t,0)} e^{-i t \Havg(t,0)}\, ,
\end{equation}
The quasienergy operator is replaced by the {\bf A}verage-{\bf E}nergy operator $\Havg(t,0){=}\sum_n \Eavg_n(t,0) \ketbra{\psi_n[0]}$~\cite{Le2022_MissingQuantumNumberFloquet}, with time-dependent but non-folded energy spectrum $\Eavg_n(t,0){=}t^{-1}\int_0^t \EKn{n}(s)\dd s$ which for $t{=}T$ equals the period-averaged Kato energies. Numerically, one can directly obtain the Kato spectrum from the lab-frame Hamiltonian by taking the diagonal expectation value of Eq.~\eqref{eq:Kato_floquet} in the Floquet states: $\EKn{n}(t){=}\bra{\psi_n[t]}H(t)\ket{\psi_n[t]}$.
Unlike the stroboscopic Floquet decomposition, the Kato decomposition~\eqref{eq:Kato_evo} naturally separates out the geometric and dynamical contributions to the time evolution. This allows us to acquire a complementary understanding of nonequilibrium phenomena: our results below indicate that the geometric phase carries the information about inherently nonequilibrium phenomena, suggesting a deeper connection between quantum geometry and nonequilibrium dynamics, while the dynamical phase is equilibrium-like. 

Practically, the uniqueness of the Kato decomposition allows us to define a unique sorting of the Floquet eigenstates according to the average-energies $\Eavg_n(T)$; this opens up the possibility of identifying the Floquet ground state and defining filling for Floquet-Bloch bands, resolving a long-standing problem of Floquet theory, and offering an alternative to the recently proposed spectral unwinding~\cite{dinc2024effective}.
Note that it is the average-energy operator $\Havg(T,0)$, and not the Kato Hamiltonian $\HK(t)$, that provides a physically meaningful spectrum; this is because $\Havg[t]\equiv\Havg(t+T,t)$, unlike $\HK(t)$, is an invariant of motion and the eigenvalues $\Eavg_n(T)$ of $\Havg(T,0)$ are manifestly independent of the phase of the drive (Floquet gauge), unlike the Kato energies~\footnote{Note that the AGP gauge (eigenstate re-phasing) is different from the Floquet gauge (phase of the drive); $\Eavg_n(T)$ is manifestly invariant w.r.t.~both gauges.}.

\begin{figure}[t]
    \centering
    \includegraphics[width=\linewidth]{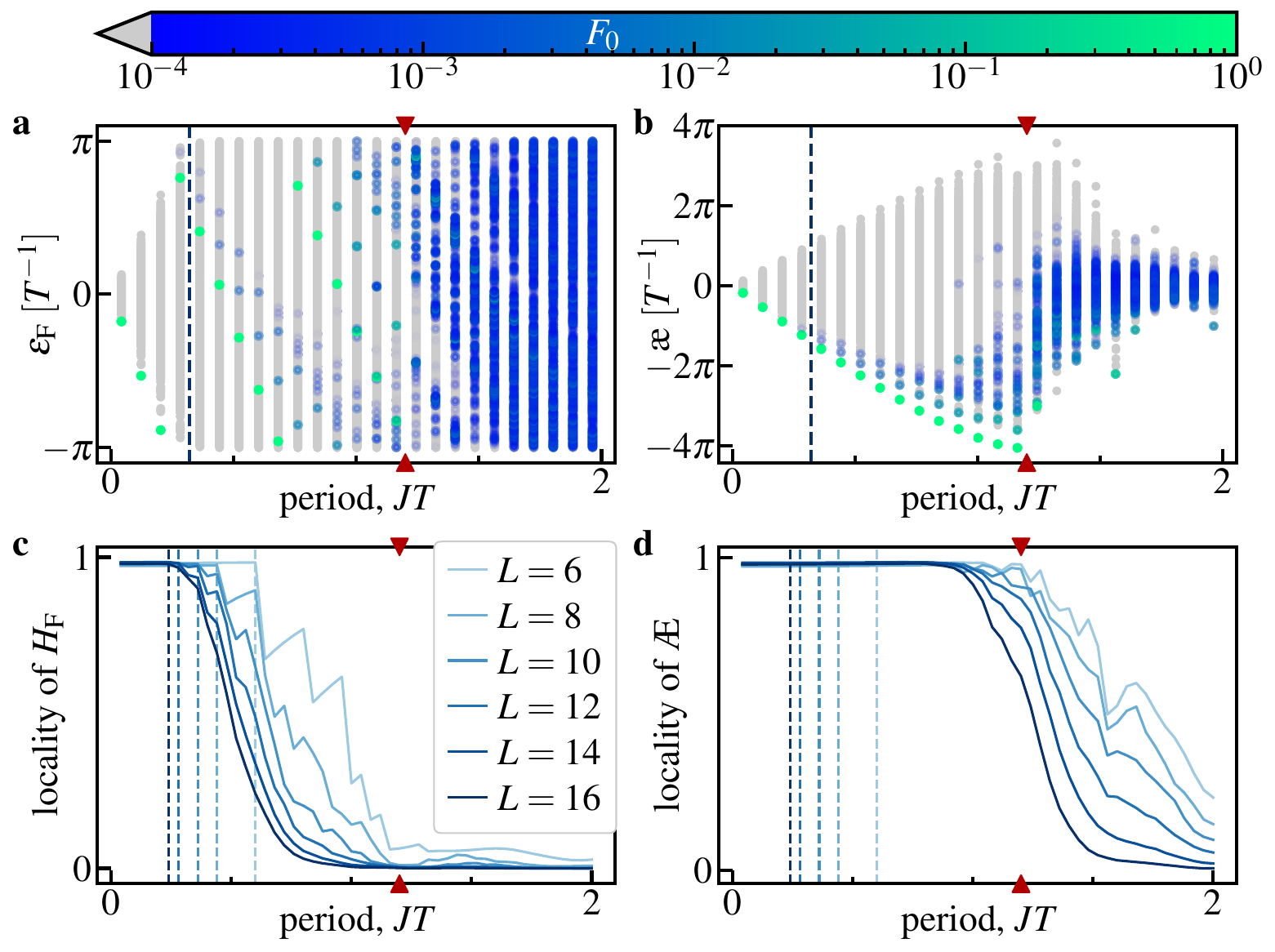}
    \caption{
    \textbf{Comparison Floquet and Kato for Floquet ground state} for kicked mixed field Ising model.
    \textbf{a}, Floquet quasienergies $\EF$, \textbf{b}, average-energies $\Eavg$
    as functions of the period $T$. Color indicates overlap $F_0{=}\abs{\braket{\psi_0}{\psi_n}}^2$ with ground state of the time-averaged Hamiltonian $H_F^{(0)}{=}H_1{+}H_2$ as indicated by color bar---values $F_0{<}10^{-4}$ are indicated as gray; dark blue dashed line indicates the period at which quasienergy folding occurs and the high-frequency expansion breaks down.
    \textbf{c} and \textbf{d}, locality of $\HF$ and $\Havg$, respectively, for different system sizes, $L{=}6,\dots,16$~(light to dark blue) measured via their relative weight on one- and two-body operators~(see also App.~\ref{app:floquetgs}):~$1$ corresponds to fully local and $0$ to fully nonlocal.
    Remarkably, the average-energy ground state remains adiabatically connected to the infinite frequency ground state in the entire regime $T{\in}[0,1.2]$, far beyond the limitations of the high-frequency expansion, until the putative heating transition occurs; red arrows indicate the transition point of Ref.~\cite{Ikeda_2024} at $JT{\approx}1.2$.
    Moreover, the average-energy operator $\Havg$ seems to undergo a locality-to-nonlocality transition at the putative heating transition; hence, $\Havg$ represents a local conserved quantity in the non-heating regime.
    We use periodic boundary conditions for $L{=}16$ spins, with $J{=}g{=}h{=}1$, all states in the zero momentum sector are shown.
    }
    \label{fig:floquetGS}
\end{figure}

As expected, in the infinite frequency limit~($\omega{\to}\infty$) one can show that $\Havg[0]{=}\HF[0]$ coincide.
Therefore, at high frequencies, the average-energy sorting of the Floquet spectrum agrees with the spectral unfolding of the inverse-frequency expansion.
However, we do not expect a simple high-frequency expansion to exist for the Kato Hamiltonian and AGP as they encode non-perturbative information about the change of eigenstates, cf.~appendix~\ref{app:HFE}; moreover, the relation between the manifestly Floquet-gauge invariant van Vleck formulation of the Floquet Hamiltonian and Eq.~\eqref{eq:Floquet_STA} is currently unclear.
Whereas pertubative expansions break down due to resonances, the average-energy sorting is well-defined at any drive frequency and in the thermodynamic limit. 
We exemplify this in Fig.~\ref{fig:floquetGS}, for a toy model recently suggested to host a Floquet ground state beyond the perturbative regime \cite{Ikeda_2024}.

\begin{figure}[t]
    \centering
    \includegraphics[width=\linewidth]{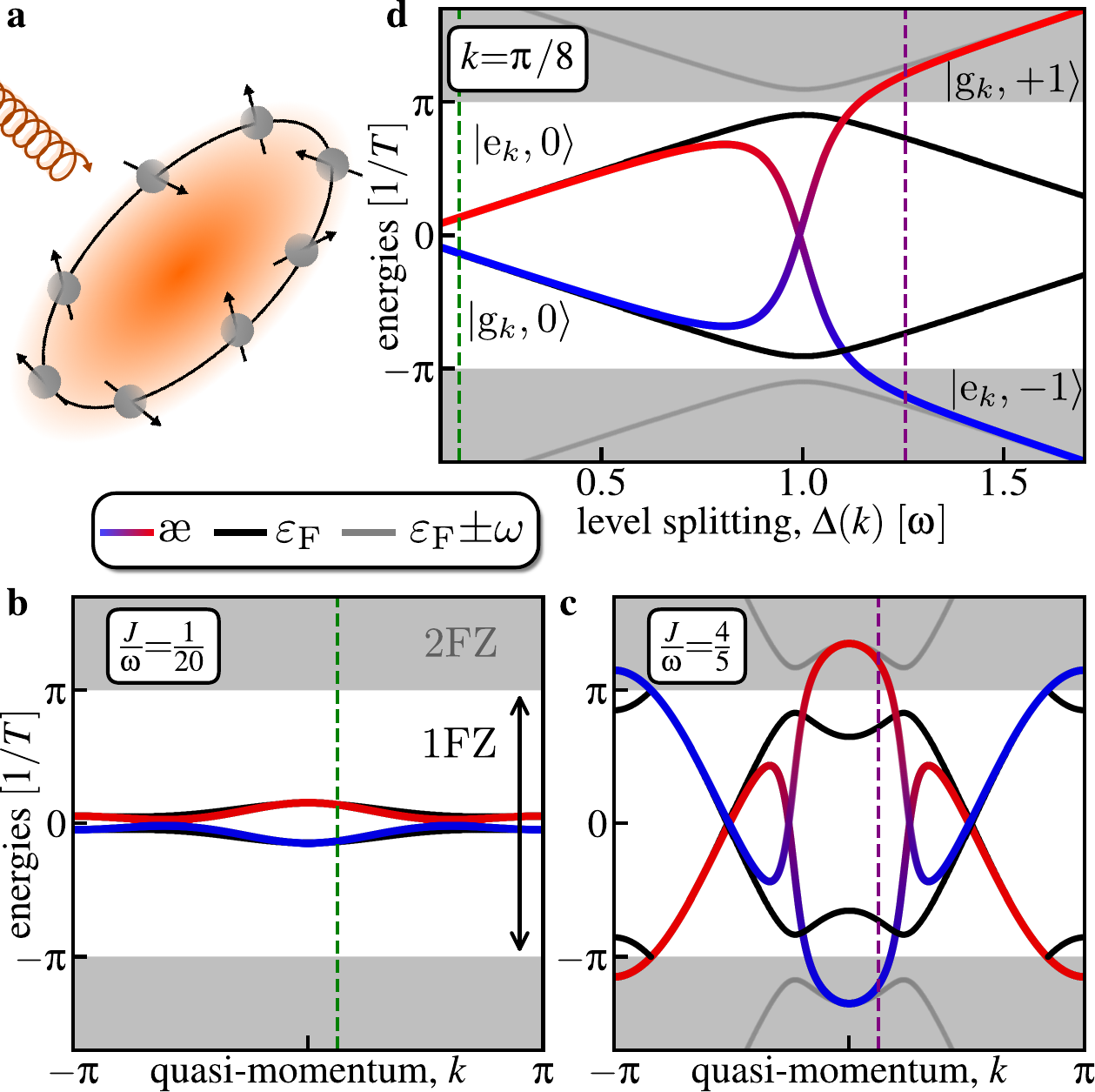}
    \caption{
    \textbf{Comparison of Floquet quasienergies and average-energies for XY model.}
    \textbf{a}, Sketch of circularly driven XY spin chain.
    \textbf{b} and \textbf{c}, momentum $k$ dependence of Floquet quasienergies $\EF(k)$~(black) and average-energies $\Eavg(k)$~(blue-red-colored) in high frequency regime and on resonance, respectively; coloring indicates overlap with average Hamiltonian~($\propto \tau^z$) ground~(blue) and excited~(red) band.
    Green and purple line indicate $(J,k)=\left(\omega/20,\pi/16\right)$ and $(J,k)=\left(\omega,\pi/16\right)$, respectively. For Floquet energies the second `Floquet Zone'~(2FZ) energies, $\EF{\pm}\omega$, are shown in gray.
    In the high-frequency regime, average-energies and quasienergies coincide. In the resonance regime, they only coincide far away from the resonance condition; near resonance the average-energies vary notably due to hybridization with the photon field.
    \textbf{d}, sweep of coupling $J$ for single momentum, $k={\pi}/{16}$, across the photon resonance near $\Delta(k){=}\omega$: notably, near the resonance, but not exactly on-resonance, cf.~appendix~\ref{app:xy}, the average-energies `hybridize' with the photon mode and after the resonance the levels absorbed/emitted a photon, indicated by ${\pm}1$ index.
    Note that all crossings of the average-energies are exact crossings.
    Other parameters are $A=2.5g$, $\omega=10g$.
    }
    \label{fig:xy}
\end{figure}

\textit{Examples.---}
Let us now use specific examples to illustrate the geometric Floquet theory and compare it to the traditional stroboscopic Floquet theory. To appreciate its significance, we focus on elucidating quantum geometric phenomena in models chosen for their nonequilibrium properties. 

First, we consider a circularly driven XY-chain in a transverse field, see Fig.~\ref{fig:xy}a,
\begin{equation}
    \label{eq:XY}
    H(t) = \frac{1}{2}\sum_{n=1}^L \left[ \left( J \sigma^+_{n+1} \sigma^-_{n} + A i e^{-i\omega t} \sigma^+_{n+1} \sigma^+_{n} + \mathrm{h.c.}\right) + \frac{g}{2} \sigma_n^z\right]\,,
\end{equation}
with periodic boundary conditions, and Pauli matrices $\sigma^\alpha_\ell$, $\alpha=x,y,z$.
Applying the Jordan-Wigner and Fourier transforms, we map the system to free fermions~(appendix~\ref{app:xy}):
\begin{equation}
    \label{eq:XY_fermion}
    \begin{aligned}
        H(t)    & \hat{=} \sum_k \boldsymbol{\psi}_k^\dagger h(k,t) \boldsymbol{\psi}_k\, ,\\
        h(k,t)  &=
            \Delta_k\tau^z + A_k\left[ \cos(\omega t) \tau^x + \sin(\omega t) \tau^y \right]\, ,
    \end{aligned}
\end{equation}
where we introduced the momentum-dependent level splitting $\Delta_k=g + J \cos(k) $ and drive amplitude $A_k=A\sin(k)$, the spinor $\boldsymbol{\psi}_k^\dagger$, and the pseudo-spin operators $\tau^\alpha=\sigma^\alpha/2$. 
The XY model~\eqref{eq:XY} and its free-fermion counterpart~\eqref{eq:XY_fermion} provide a paradigmatic example of quantum phase transitions, conventional superconductivity, and topological insulators~\cite{Barouch_etal_XY_1971,Fradkin1978_XY,Read2000_Freefermion}, respectively.
Here, we make use of its exact solvability to study the physical properties of the Kato formalism, and contrast the latter with the stroboscopic Floquet approach; while the model is exactly solvable, it exhibits key properties of driven systems like photon resonances.

Since the free-fermion representation describes independent two-level systems, one for each quasi-momentum $k$, we can directly obtain the Floquet Hamiltonian and AGP, cf.~appendix~\ref{app:xy},
\begin{equation}
    \begin{aligned}
        h_{\F,k}[t] &= \pqty{\epsilon_k - \omega} \left\{ \frac{\Delta_k-\omega}{\epsilon_k} \tau^z + \frac{A_k}{\epsilon_k} \left[ \cos(\omega t) \tau^x + \sin(\omega t) \tau^y \right]  \right\} \, ,\\
        a_{\F,k}(t) &= \omega\Bqty{ \pqty{1 + \frac{\Delta_k-\omega}{\epsilon_k}} \tau^z + \frac{A_k}{\epsilon_k} \left[ \cos(\omega t) \tau^x + \sin(\omega t) \tau^y \right]}\, \,,
    \end{aligned}
\end{equation}
with $\epsilon_k^2=(\Delta_k-\omega)^2+A_k^2$.
Using the eigenstate projectors, $\Pi_{\pm}= \pqty{\identity \pm h_{\F,k}[t]}/2(\epsilon_k-\omega)$, and $a_{\F,k}(t)=\frac{1}{2}\sum_n\comm{\Pi_n}{i\partial_t \Pi_n}$, we can also readily compute the Kato Hamiltonian and gauge potential
\begin{equation}
    \begin{aligned}
        h_{\K,k}(t) &= \frac{\epsilon^2_{\K,k}}{\epsilon_k} \left\{ \frac{\Delta_k-\omega}{\epsilon_k} \tau^z + \frac{A_k}{\epsilon_k} \left[ \cos(\omega t) \tau^x + \sin(\omega t) \tau^y \right]  \right\} \, ,\\
        a_{\K,k}(t) &= \frac{A_k\omega}{\epsilon_k} \left\{ \frac{A_k}{\epsilon_k} \tau^z -  \frac{\Delta_k-\omega}{\epsilon_k}\left[ \cos(\omega t) \tau^x + \sin(\omega t) \tau^y \right]  \right\} \, ,
    \end{aligned}
\end{equation}
with $\epsilon^2_{\K,k}=\left(\Delta_k-\omega\right)\Delta_k+A_k^2$. Notably, in this specific case $\Havg_{k}=h_{\K,k}$ since the eigenvalues of $h_{\K,k}$ are independent of the phase of the drive.
Inverting the Jordan-Wigner transform, we find that both $\HF$ and $\HK$ have exponentially localized support in the original real-space spin representation.

Clearly, $h(k,t)=h_{\F,k}+a_{\F,k}=h_{\K,k}+a_{\K,k}$, and in the high-frequency limit~($\omega\to\infty$) Kato and Floquet agree, $h_{\F,k},h_{\K,k}\to \Delta_k \tau^z$, see also Fig.~\ref{fig:xy}b. However, note that $a_{\K,k}\to A_k \left[ \cos(\omega t) \tau^x + \sin(\omega t) \tau^y \right]$ is not a valid Kato AGP for $h_{\K,k}\mid_{\omega=\infty}$, pointing at a difficulty with defining a high-frequency expansion for the Kato objects (see appendix~\ref{app:HFE}).

By contrast, if the drive is resonant with the energy of the system, $\epsilon_k \approx \omega$ for some $k$, the Floquet quasienergies and average-energies can differ substantially, see Fig.~\ref{fig:xy}c, with the average-energy bands intertwining near resonance. In contrast to Floquet quasienergies $\EF$ which can always be chosen to lie within the first Floquet zone, $-\omega/2<\EF \leq \omega/2$, the average-energies can exceed this window, as exemplified near $k\approx0$ in Fig.~\ref{fig:xy}c.

Taking a closer look at a specific momentum value and sweeping from the high-frequency regime past the resonance point, see Fig.~\ref{fig:xy}d, the intertwining of average-energy bands near resonance becomes clear: when the average-energies become resonant with the drive, hybridization with the driving field can occur, and the system can eventually emit and absorb a drive quantum $\omega$, leading to an interchange of ground and excited state, respectively. Note that the resulting crossing of average-energy levels is an exact crossing; the position of the crossing is close to but does not coincide with the position of the resonance~(instead the resonance coincides with a maximum in geometric phase, $\abs{\gamma}=\pi$).
Therefore, the average-energies properly take into account the energy balance due to the absorption and emission of drive quanta [Fig.~\ref{fig:xy}d], even though there is no photon degree of freedom in the system.
Hence, one can uniquely define a physically meaningful Floquet ground state via the average-energies.

Next, we consider the non-integrable kicked Ising model described by unitary evolution $U_\F {=} U_1 U_2 U_1$ where $U_m{=}\exp(-i mTH_m/4 )$, $H_1{=}{-}\sum_{n=1}^L { (J/2) \sigma^z_{n}\sigma^z_{n+1} {+} h \sigma_n^z}$, and $H_2{=}{-}g\sum_{n=1}^L \sigma_n^x$.
Numerical studies~\cite{Heyl2019_LocalizationTrotter,Ikeda_2024} suggest that this model hosts a low-entropy state that, surprisingly, avoids heating up to a finite period~($JT{\approx}1.2$) beyond the regime of validity of the high-frequency expansion.
A ground-state-like character of this state is anticipated since it retains finite overlap with the ground state at infinite frequency, deep inside the non-heating regime, see Fig.~\ref{fig:floquetGS}a 

We now provide a rigorous argument for why this state should be considered the Floquet ground state deep into the regime $JT{\sim}1$ where quasienergy folding is inevitable.
Indeed, this state emerges as the lowest-energy state of the unique average-energy operator, see Fig.~\ref{fig:floquetGS}b.
Moreover, near the putative phase transition this state undergoes photon resonances; as a result, the many-body bandwidth first grows as $JT{\to}1.2$, and then implodes making the critical point (red triangle) also apparent in the spectrum; finite-size scaling is shown in appendix~\ref{app:floquetgs}.
Therefore, the spectrum $\Eavg(T)$ is well-suited to identify the Floquet ground state, providing a reliable tool to seamlessly connect the infinite frequency behavior to finite frequencies.

Figure~\ref{fig:floquetGS}c,d shows the relative weight of one- and two-body terms in $\HF[0]$ and $\Havg(T,0)$, respectively: unity corresponds to a two-body operator, and zero indicates spatial nonlocality~(see App.~\ref{app:floquetgs} for details). Notice that $\HF[0]$ becomes nonlocal when quasienergy folding occurs as indicated by the curves shifting to the left with increasing system size; by contrast, $\Havg[0]$ remains local all the way up to $JT{\to}1.2$ almost independent of the system size.
Since $\Havg$ is an invariant of motion, the Floquet system admits a local conservation law.
The locality and ground state gap of $\Havg$ imply the existence of low-entanglement Floquet states.
The finite average-energy gap persisting throughout the entire non-heating regime~($0 < JT < 1.2$) guarantees that this Floquet ground state is adiabatically connected to the ground state of the (static) mixed field Ising model; this allows for efficient state preparation in experiments, for instance, by adiabatically decreasing the frequency of the drive~\cite{Schindler2024_FCD}.
The nonlocality of $\HF$ in many-body systems can be explained by the ambiguity of defining $\HF{\mapsto}\HF{+}m\omega\ketbra{\psi_k}$ ($m{\in}\mathbb{Z})$ due to shifts in the quasienergy of individual Floquet states, as the Floquet eigenstate projector $\ketbra{\psi_k}$ is highly nonlocal.
Because the average-energy is sorted, this argument does not apply to $\Havg(T,0)$.
This behavior highlights the utility of the geometric formulation of Floquet theory in identifying local physical generators of the dynamics.

More generally, for local lab-frame Hamiltonians $H(t)$ Eq.~\eqref{eq:Floquet_STA} implies that the non-locality of the Floquet Hamiltonian $\HF[t]$ in generic Floquet many-body systems~\cite{Bukov_2015_general_HFE,Kuwahara2016_prethermalization,Ho2023_prethermalization} can be traced back to the non-locality of the AGP $\A_F(t)$~\cite{Kolodrubetz2017_GeometryReview,Sels2017_LCD} (and vice-versa). Therefore, whenever $H(t)$ is a local Hamiltonian, the nonlocality of the Floquet Hamiltonian is directly related to the nonlocality of the AGP.
Similarly, Eq.~\eqref{eq:Kato_floquet} implies that the nonlocal character of the Kato Hamiltonian $\HK(t)$ is inherited by that of $\A_K(t)$. 
A thorough comparison between the locality properties of $\Havg(t,t_0)$ ($\HK(t)$) and $\HF[t]$ is an interesting open question for future investigation.

Likewise, the average-energy spectrum is also instrumental in determining the correct filling of Floquet energy bands for high (but finite) frequencies, e.g., in the case of Floquet-engineered Haldane models~\cite{Haldane1988,Jotzu2014_Haldane,Quelle2017_Haldane,Mishra2018_Haldane,Mishra2018_Haldane}.
In this case, we find that the average-energy spectrum agrees with the results of the high-frequency expansion, and $\Eavg(T,0)$ provides a rigorous procedure to define energy states unambiguously at any frequency.

\begin{figure}[t]
    \centering
    \includegraphics[width=\linewidth]{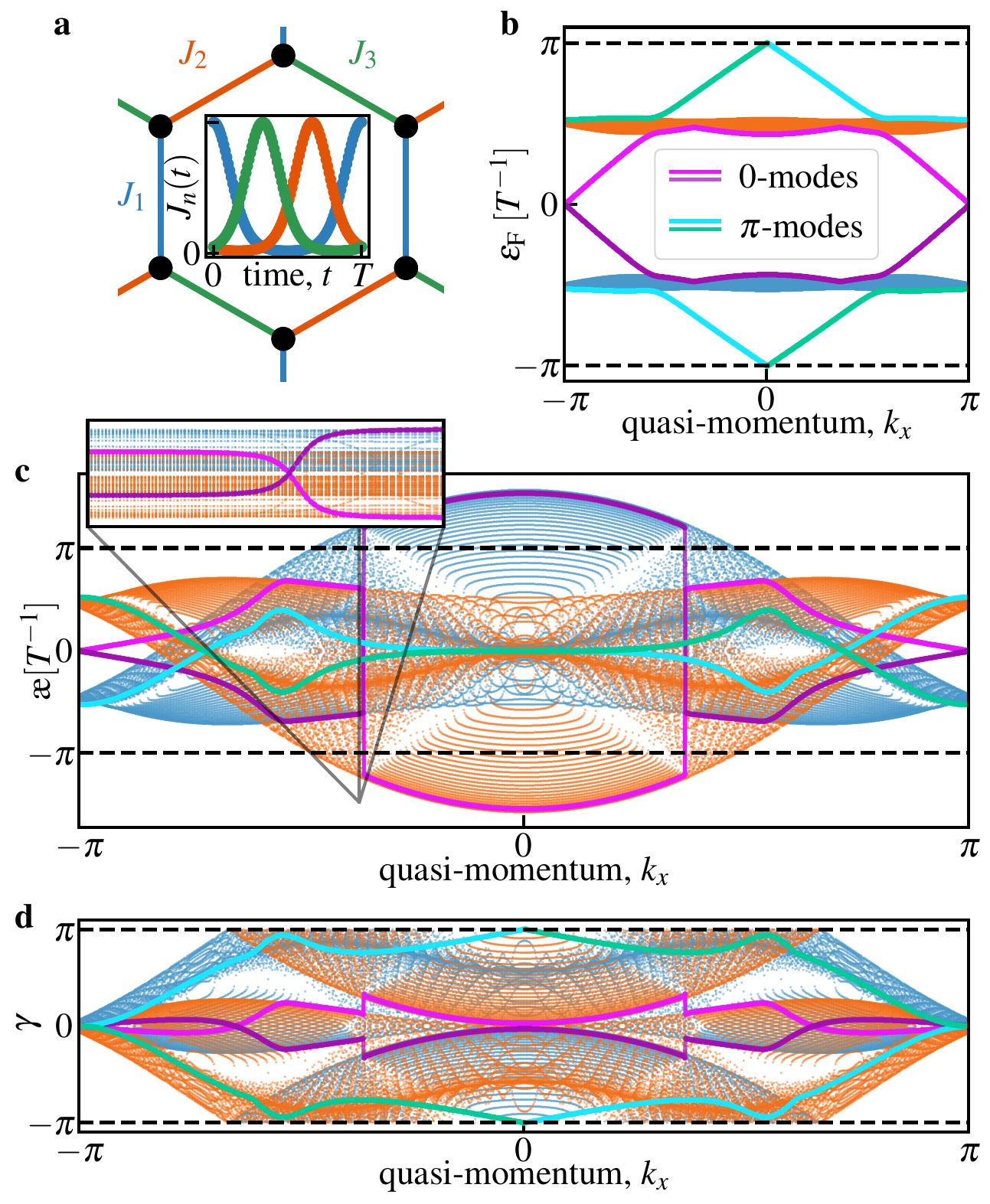}
    \caption{
        \textbf{Anomalous Floquet topological insulator}, comparison geometric vs.~stroboscopic Floquet theory.
        \textbf{a}, time-dependence of the three distinct hopping amplitudes, $J_{1,2,3}$, in the unit cell;
        \textbf{b}, quasienergies $\EF(k_x)$ show two distinct flat bands with two $0$~(pink, purple) and two $\pi$~(cyan, green) edge modes.
        \textbf{c}, average energies $\Eavg(k_x)$.
        The two bands become intertwined due to photon-resonances occurring throughout the Brillouin zone, and the edge states are no longer distinguishable from the bulk states in the average energy spectrum.
        \textit{Inset} shows zoom into photon resonance of $0$-modes, revealing their smooth dependence on $k_x$.
        \textbf{d}, (folded) Berry phases $\gamma(k_x)$ show characteristic winding behavior for the $\pi$-modes.
        Calculations are performed on a cylindrical geometry, circular in $x$-direction, on system size $(L_x,L_y)=(500,50)$; zoom-in window $-0.3695<k_x/\pi<-0.37$ for inset contains $100$ $k_x$ points.
    }
    \label{fig:AFTI}
\end{figure}

\textit{Inherently nonequilibrium examples.---}%
Intriguingly, periodically driven systems can also lead to physics without static counterpart; to describe such phenomena considering the Floquet Hamiltonian alone is often insufficient.
In the following, we discuss two different types of inherently nonequilibrium phenomena to explore the geometric Floquet decomposition. Specifically, we focus on anomalous Floquet topological insulators~(AFTI) and discrete time crystals~(DTCs), as paradigmatic examples for drive-induced topological order and time-translation-symmetry breaking.

As an AFTI, we consider the continuously driven free fermion model on a honeycomb lattice~\cite{Dutta2024_AFTI}, experimentally realized in Ref.~\cite{Wintersperger2020_Anomalous}, where hopping amplitudes are varied in a spatially homogeneous but chiral, time-periodic way, cf.~Fig.~\ref{fig:AFTI}a (details in appendix~\ref{app:AFTI}).
Notably, in AFTIs Floquet bands are insufficient to classify bulk topology; instead, the micromotion operator reveals the non-trivial topology of the system~\cite{Rudner2013_AnomalousFloquetEdgeState,Nathan2015_AnomalousFloquetPrediction}.
As a consequence, on a cylinder, this model hosts \textit{anomalous} edge states, which are topologically protected despite the topologically-trivial flat bulk bands~\cite{Rudner2013_AnomalousFloquetEdgeState,Nathan2015_AnomalousFloquetPrediction,Nathan2019_AnomalousFloquet,Nathan2021_anomalous,Wintersperger2020_Anomalous}. In particular, this manifests in the appearance of a $\pi$-mode, i.e., an edge state connecting Floquet bands in different Floquet zones, cf.~Fig~\ref{fig:AFTI}b.

Due to the absence of the quasienergy folding paradigm in the Kato decomposition, the same cannot happen for the average-energy bands, see Fig.~\ref{fig:AFTI}c.
Instead, the two $\pi$-modes cross at zero energy and become indistinguishable from the conventional $0$-edge modes.
To recover the nonequilibrium behavior of the $\pi$-modes the Berry phase~(Fig.~\ref{fig:AFTI}d) must be considered as well. Analogous to the quasienergy spectrum, the Berry phase spectrum shows the characteristic winding of the $\pi$-modes, distinguishing them from the $0$-modes.
Moreover, it has recently been shown that the anomalous winding number $\mathcal{W}_\mathrm{A}$, determining the number of anomalous edge states, can be obtained from the linear magnetic response of the bulk geometric phases: $\mathcal{W}_\mathrm{A}=-(\Phi_0/2\pi) \partial_\Phi \sum_n \gamma_n$ with normal flux quantum $\Phi_0=hc/e$~\cite{gavensky2024stredaformulafloquetsystems}.

Noticeably, the two $\Eavg$-bands are no longer flat and overlap due to the presence of photon resonances [cf.~discussion on XY model]; this raises the intriguing question as to which band is filled and which remains empty in a half-filled system. While the average-energy and Floquet spectra can deviate substantially, band topology as measured by Chern numbers is determined by eigenstates which are the same for $\HF$ and $\Havg$.
We emphasize that no avoided crossings occur so that the unique sorting between the upper and lower $\Eavg$-bands can be recovered starting from $k_x{=}0$ and tracing the states adiabatically.

The multitude of sharp photon resonances (e.g, Fig.~\ref{fig:AFTI}c inset) leads to a ``chaotic'' regime in the $\Eavg$-bands~($\abs{k_x}{<}\pi/2$ in Fig.~\ref{fig:AFTI}c) where a small change in $k_x$ leads to large changes in the spectrum; this behavior is even more pronounced in multiphoton resonances and occurs in other AFTI models, cf.~appendix~\ref{app:AFTI}. Hence, we anticipate that the $\Eavg$-spectrum can be used to distinguish chaotic from non-chaotic Floquet states, as captured by the sensitivity of eigenstates to changes in $k_x$ measured by the AGP~\cite{pandey2020adiabatic}.
This photon-resonance-induced ``chaos'' in the average-energy spectrum can serve as an indicator for heating in many-body Floquet systems, unifying previous attempts to detect the onset of heating via chaos~\cite{pandey2020adiabatic,bhattacharjee2024_AGPheating} and Floquet resonances~\cite{Bukov2016_resonances}.

In short, we find that the geometric phase is required to distinguish anomalous Floquet from conventional static topological edge states, which underscores the fundamental role of geometry in AFTIs; the average-energy spectrum alone is insufficient to tell these states apart.
We expect similar results to hold for other Floquet-induced topological edge states, such as $\pi$-Majorana modes which generalize to interacting systems~\cite{Nathan2021_anomalous,yates2019almost}.

\begin{figure}[t]
    \centering
    \includegraphics[width=\linewidth]{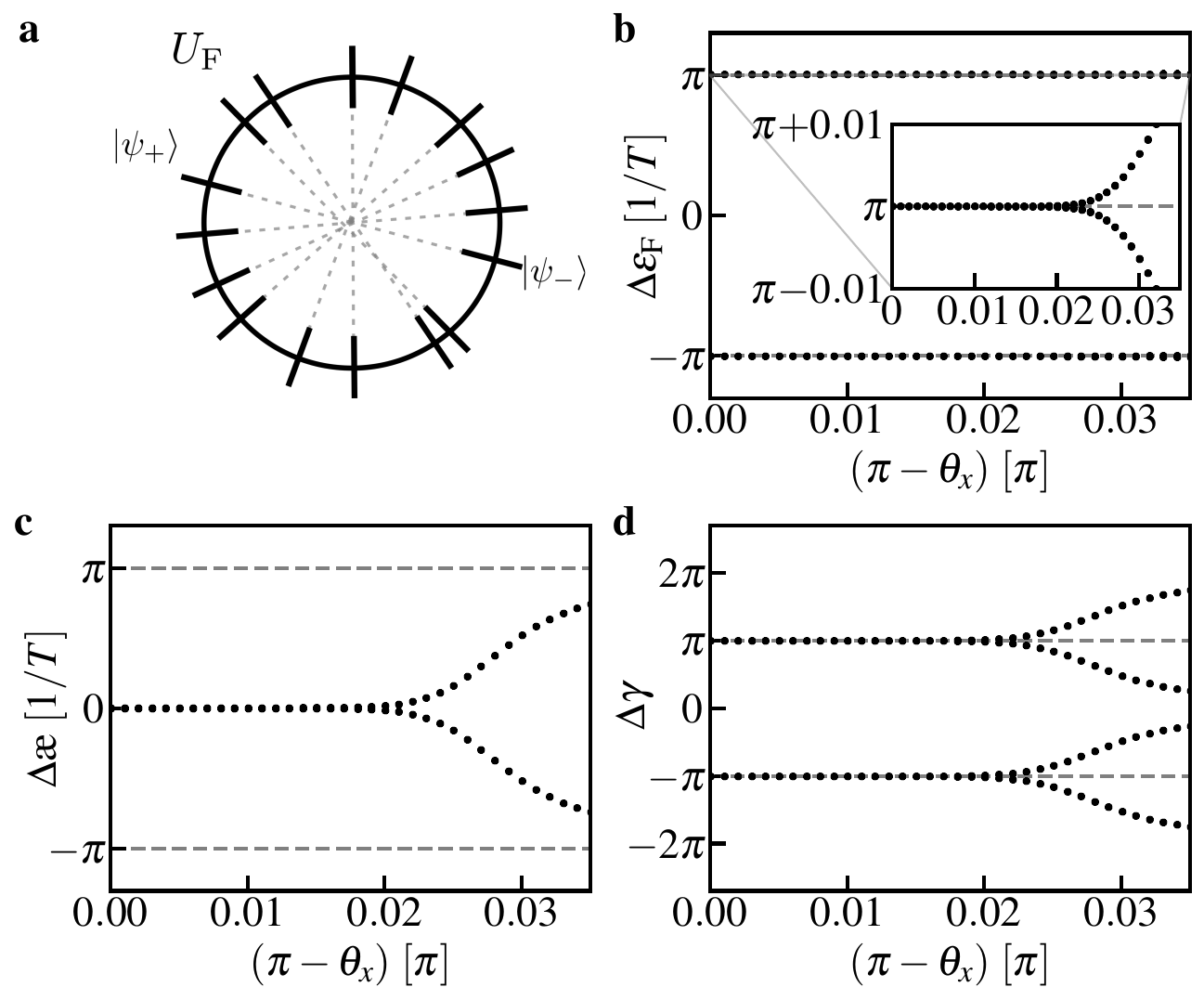}
    \caption{
    \textbf{Quantum geometric analysis of discrete time crystal} in kicked Ising spin chain.
    \textbf{a}, sketch of $\pi$-pairing of spin-inversion even ($+$) and odd~($-$) sector.
    \textbf{b}, difference in Floquet energies $\Delta \EF=\pqty{\EFn{+}-\EFn{-}}$ between positive~($\EFn{+}$) and negative~($\EFn{-}$) spin-inversion sector as a function of kick-angle $\theta_x$. 
    The DTC behavior is evident in the $\pi$-pairing of the quasienergies that survives away from the fine-tuning point $\theta_x=\pi$.
    \textit{Inset:} zoom into $\Delta\EF T = \pi\pm0.01$ region indicates breakdown of $\pi$-splitting around $\pi-\theta_x\approx 0.02\pi$.
    \textbf{c} and \textbf{d}, average-energy difference $\Delta \Eavg=\Eavg_{+}-\Eavg_{-}$ and geometric phase difference $\Delta \gamma=\gamma_+-\gamma_-$ reveal that the DTC behaviour is of purely geometric origin, i.e., the quasienergy difference is determined by the geometric phases $\Delta \EF T =- \Delta \gamma$ in the DTC-regime~[$(\pi-\theta_x) \leq 0.02\pi$].
    We consider an open spin-chain of length $L{=}14$ with disordered nearest neighbor couplings $J_i{=}J(1{+}\eta_i)$ with random uniform $\eta_i{\in}\bqty{-0.5,\,0.5}$ and period $JT=0.05$.
    Eigenstates $\psi_{n,\alpha}[\theta_x]$ are computed for each parity sector~($\alpha=+,-$) individually; pairs are identified via the exact $\pi$-splitting at $\theta_x=\pi$, and then adiabatically followed for different $\theta_x$ values.
    }
    \label{fig:dtc}
\end{figure}

Finally, we consider discrete time crystals~(DTCs)~\cite{Khemani2016_DTC,Yao2017_DTCRigidity,Else2020_DTCReview,Zaletel2023_DTCReview} -- nonequilibrium ordered states without equilibrium counterpart. For concreteness, consider the Floquet operator $U_\F(\theta_x)=e^{-i T H_\mathrm{int}} e^{-i \theta_x H_x}$ with Ising interactions $H_\mathrm{int}=\sum_{n} J_{n} \sigma^z_n \sigma^z_{n+1}$ and on-site field $H_x=\sum_n \sigma^x_n/2$; importantly, the Ising interactions admit a $\mathbb{Z}_2$-symmetry $\comm{H_\mathrm{int}}{\mathcal{P}_x}=0$ with $\mathcal{P}_x=\exp(-i\pi H_x)$ and $\mathcal{P}_x^2=\identity$. 
In the $\mathbb{Z}_2$-ordered phase, this symmetry is spontaneously broken, i.e., there exist eigenstates $\ket{\psi}$ of $H_\mathrm{int}$ that transform non-trivially under the symmetry $\mathcal{P}_x$; this induces spatiotemporal~(DTC) order for $\theta_x{\approx}\pi$.
A hallmark feature of DTC order is the robustness towards $\theta_x$-perturbations originating from the $\pi$-gap between the odd- and even-parity states $\ket{\psi_\pm}\propto\ket{\psi}{\pm}\mathcal{P}_x\ket{\psi}$~\footnote{This is a result of $\ket{\psi_\pm}$ being eigenstates of both $\mathcal{P}_x$ and $H_\mathrm{int}$ with the same energy but opposite parity $\mathcal{P}_x\ket{\psi_\pm}=\pm \ket{\psi_\pm}$.}.
By contrast, these two states remain exactly degenerate in the average-energy; indeed, $\Havg[0]{=}H_\mathrm{int}$, since $\bra{\psi_\pm}H_\mathrm{int}\ket{\psi_\pm}{=}\Eavg{=}\EK$ and $\bra{\psi_\pm}H_x\ket{\psi_\pm}{=}0$.
Therefore, the quasienergy splitting in the Floquet spectrum is of purely geometric origin.
If DTC order survives away from the fine-tuning point $\theta_x{=}\pi$, we expect similar arguments to hold, since then the Floquet unitary can be generically expressed as $U_\F=\mathcal{U}^\dagger \mathcal{P}_x e^{-iDT}\mathcal{U}$ with a time-independent change of basis $\mathcal{U}$ and a quasi-local effective Hamiltonian $D{=}\sum_{n,m} J_{nm}\sigma^z_n \sigma^z_m {+} \sum_{nmpq} K_{nmpq} \sigma^z_n \sigma^z_m\sigma^z_p \sigma^z_q+{\cdots}\;$~\cite{Else2017_DTCRobust}. 

Numerical simulations on a disordered kicked chain of $L{=}14$ spins support our analytical analysis, see Fig~\ref{fig:dtc}:
in a finite parameter regime around the fine-tuned point, $\theta_x{=}\pi$, we observe $\pi$-splitting between quasienergies accompanied by the same splitting in geometric phases and degenerate average energies.
Further away from the fine-tuned point, $(\pi-\theta_x){>}0.02\pi$, the average energies~(Fig~\ref{fig:dtc}c) are no longer degenerate, clearly indicating a transition away from DTC order; the variation in quasienergy~(Fig~\ref{fig:dtc}b) is noticeably less pronounced.
Therefore, similar to the heating transition above, the average-energy spectrum may serve as a susceptible indicator for the transition from an ordered to a disordered state.
In summary, we demonstrated that true DTC order is of purely geometric origin, while the average energies remain degenerate.

\textit{Discussion \& Outlook.---}%
We have provided an alternative formulation of Floquet theory based on concepts from quantum geometry, such as gauge potentials and geometric phases. The key insight is the identity~\eqref{eq:Floquet_STA} demonstrating that the lab frame Hamiltonian $H(t)$ generates counterdiabatic driving for the Floquet states.
Its formulation in the parallel-transport gauge~\eqref{eq:Kato_floquet} leads to a geometric decomposition of the evolution operator~\eqref{eq:Kato_evo}, where the dynamics naturally separates into the Wilson line operator $\mathcal{W}(t,t_0)$ that accumulates geometric phases, and a dynamical part generated by the unique average-energy operator $\Havg(t,t_0)$.
The geometric phase $\gamma(T)$ and average-energies $\Eavg(T)$ can be obtained from the response of the quasienergy to changes in the period, $\gamma(T)=T^2 \partial_T \EF$~\cite{Fainshtein1978some} and $\Eavg=\partial_T(T\EF(T))$~\cite{GRIFONI1998contemporary}, respectively.
Moreover, there exist protocols to measure the Berry phase $\gamma(T)$ in experiments for few-level systems and weakly-interacting Bose-Einstein condensates with ultracold atoms~\cite{atala2013direct,duca2015aharonov,bruggenjurgen2024phase}, and using quantum computers for many-body states~\cite{iqbal2024non}.
It is also feasible to measure the average-energy, $\Eavg{=}T^{-1}\int^T_0 \langle\psi_n[t]|H(t)|\psi_n[t]\rangle\dd t$, by adiabatically preparing the Floquet states $|\psi_n[t]\rangle$~\cite{Schindler2024_FCD}, and measuring the time-resolved expectation value of the lab-frame Hamiltonian~\cite{impertro2024local}.
Using the relation $\EF{=}\Eavg(T,0){+}T^{-1}\gamma(T)$ this provides a new way to map out the quasienergy spectrum, and shows that the central quantities of geometric Floquet theory are experimentally accessible.

Conceptually, a key result of geometric Floquet theory is the inherent geometric origin of nonequilibrium phenomena, encoded in the Berry phases of the Floquet states.  
For anomalous Floquet topological insulators, we show that the geometric phases carry the relevant information about nonequilibrium properties since Floquet ($\pi$) edge modes are indistinguishable from ordinary ($0$) edge modes in the average-energy spectrum; 
in discrete time crystals, the characteristic $\pi$-splitting between quasienergy states manifests only in the geometric phase, as parity-odd and -even states have degenerate average-energy.
Therefore, geometric Floquet theory provides a natural decomposition of the evolution into dynamical equilibrium-like properties and geometric nonequilibrium properties.

We have also presented a derivation of Floquet theory from the adiabatic theorem. 
In the small-frequency regime, the Kato decomposition~\eqref{eq:Kato_floquet} provides the proper framework to understand adiabatic evolution. 
This enabled us to interpret quasienergy folding as a consequence of partial gauge fixing, occurring as a result of a broken gauge group of the adiabatic gauge potential $U(1){\mapsto}\mathbb{Z}$.  
Instead, based on the unique parallel-transport gauge the Kato decomposition~\eqref{eq:Kato_floquet} unambiguously sorts the quasienergy spectrum and identifies the Floquet ground state as the lowest-energy state of the average-energy operator $\Havg(T,0)$. 
In the infinite-frequency limit, this sorting agrees with the high-frequency expansion; remarkably, the Kato decomposition is non-perturbative and applies to all frequencies. The average-energy spectrum properly accounts for the photon absorption/emission counting when passing through photon resonances, as we demonstrate on the exactly solvable XY model.
Moreover, it serves as an indicator of nonequilibrium phase transitions, capturing the occurrence of both spatiotemporal symmetry-breaking and heating transitions, and can be used to identify new types of nonequilibrium order.
Remarkably, $\Havg(T,0)$ remains local in regimes where quasienergy unfolding renders $\HF$ nonlocal.  
While a comprehensive understanding of the implications of the Floquet ground state is lacking at this stage, intriguing possibilities remain in defining various types of order or preparing a thermally occupied extensive spectrum of $\Havg(T,0)$. 

In practice, geometric Floquet theory paves the way to finding approximations to the Floquet Hamiltonian and the Floquet ground state for large quantum systems.
Whenever the lab-frame Hamiltonian $H(t)$ is known, computing the Kato AGP $\AK$ immediately provides the Kato Hamiltonian $\HK(t)$. Whereas in generic systems this is rarely feasible exactly, there exists a quadratic variational principle for the AGP~\cite{Sels2017_LCD}; this allows one to construct efficient non-perturbative approximations of both $\AK$ and $\HK$ that are not limited to small system sizes~\cite{kim2024variational,mckeever2024towards}. 
Therefore, our work enables the direct application of a large body of techniques and methods from the fields of counterdiabatic driving and shortcuts to adiabaticity, to Floquet systems, establishing a bridge between seemingly unrelated areas of nonequilibrium dynamics. 
This unifying perspective explains previous studies using CD techniques to achieve dynamical freezing~\cite{gangopadhay2024_DynamicalFreezingCD} or engineer Floquet scar states~\cite{Ljubotina2022_MPS_CDScars}. 
Crucially, the plethora of numerical algorithms developed for CD driving~\cite{Saberi2014_NonlocalAGP,Petiziol2018_FloquetFastAdiabatic,Sels2017_LCD,Ieva2023_COLD,morawetz2024efficient}, can now be directly deployed for Floquet engineering.

Interestingly, our framework suggests a natural classification for distinctive families of periodic drives, Fig.~\ref{fig:introduction}e. 
According to the Floquet decomposition~\eqref{eq:Floquet_STA}, the extreme cases are: 
(i) \textit{equilibrium `drives'}, defined by $\A_\F(t){\equiv}0$, which implies $H(t){=}H$ is static;
(ii) \textit{pure-micromotion} drives obey $H(t){=}\A_\F(t)$; they have a vanishing Floquet Hamiltonian with dynamics entirely determined by the periodic micromotion $U(t,0){=}P(t)$. These are instances of Floquet systems whose evolution is periodic itself with the same period as the drive.
Examples from the literature are scarce and include Floquet flat bands~\cite{banerjee2024exact} and some anomalous Floquet topological insulators~\cite{Rudner2013_AnomalousFloquetEdgeState}. Pure-micromotion drives do not lead to energy obsorption, and hence cannot heat up; in a sense, this family of drives represents minimally nonequilibrium systems.   
According to the Kato decomposition~\eqref{eq:Kato_floquet}, we distinguish:
(iii) \textit{flat drives} with $H(t){=}\HK(t)$ and $\A_\K(t){\equiv}0$: this implies that the Floquet eigenprojectors are constant in time, $\partial_t \left(\ketbra{\psi_n(t)}\right){=}0$, and hence only the Kato energies change in time.
Flat drives admit a local Floquet Hamiltonian $\HF[0]{=}\Havg(T,0){=}T^{-1}\int_0^T H(t)\mathrm{d}t$ that has an unambiguous ground state. Therefore, flat-driven systems are equilibrium-like: they do not heat up at any drive frequency. An example is the Wannier-Stark model on a closed chain~\cite{kolovsky2003floquet}.
Last, %
(iv) \textit{pure-geometric drives} are defined by $H(t){=}\AK(t)$: the name derives from the evolution consisting solely of the geometric phases. Since $\HK(t){\equiv}0$, the $\Eavg$-spectrum vanishes identically, and one cannot sort the Floquet states; hence, pure-geometric drives do not possess a Floquet ground state -- they are maximally nonequilibrium.  
These distinctive families of periodic drives overlap and are not exhaustive; they identify special `corners' in the space of periodic drives, Fig.~\ref{fig:introduction}e, which can help us construct minimal models to understand the general behavior of Floquet systems. 

Finally, in the big picture of nonequilibrium physics, our work explicitly demonstrates that adiabaticity requires neither a Hamiltonian nor the concept of energy; it is an intrinsic property of the geometry of the state manifold. This may appear counterintuitive, as adiabaticity is usually introduced via slow variations of a parameter in a Hamiltonian. However, Floquet dynamics is not generated by a static local Hamiltonian; the physically meaningful object is the unitary operator over one drive cycle. Nonetheless, it is possible to generalize the notion of adiabaticity to periodically driven systems using the adiabatic gauge potential which generates adiabatic evolution of Floquet states~\cite{Weinberg_2017_adiabatic_Floquet,grattan2023exponential,Schindler2024_FCD}. We believe that these concepts are straightforward to define in quasi-periodically driven systems~\cite{Dumitrescu2018_QuasiPeriodic,else2020longlived,long2022manybody} and random structured drives~\cite{Hongzheng2021_RMD}, and likely also deeper into the nonequilibrium regime.

\textit{Acknowledgments.---}
We are particularly grateful to David M.~Long for a critical proofreading of a previous version of this manuscript, and to Zuo Wang for providing useful feedback.
We also thank J.~Budich, M.~Heyl, and A.~Polkovnikov for fruitful discussions. 
Funded by the European Union (ERC, QuSimCtrl, 101113633). Views and opinions expressed are however those of the authors only and do not necessarily reflect those of the European Union or the European Research Council Executive Agency. Neither the European Union nor the granting authority can be held responsible for them.
Numerical simulations were performed on the MPIPKS HPC cluster.

\textit{Data availability statement}
The data associated with this manuscript is available under DOI:10.5281/zenodo.16377692~\cite{schindler2025geometric}.

\appendix

\section{Proof of existence and uniqueness of Kato decomposition}
\label{app:proof_kato}

The results in this manuscript hinge on the existence of a unique Kato decomposition, $H(t)=H_\K(t)+\A_\K(t)$ for any periodic Hamiltonian $H(t)$.
Here, we provide two different proofs: first, we provide a simple proof relying on the validity of Floquet's theorem;
second, to complete the geometric proof of Floquet's theorem we provide a technical self-consistent proof that does not rely on Floquet theory.

\subsection{Proof based on Floquet's theorem}

Let us begin by reiterating Floquet's theorem, Eq.~\eqref{eq:Floquet},
\begin{equation*}
    \HF = P^\dagger(t) H(t) P(t)- P^\dagger(t)i\partial_t P(t),
\end{equation*}
with Floquet Hamiltonian $\HF=\HF[0]$, $\exp(-iT\HF)=\mathcal{T}\exp(-i\int_0^T H(t)\dd t)=U(T,0)$, and periodic micromotion operator $P(t)=U(t,0)\exp(+it\HF)=P(t+T)$.
Unitarily transforming Eq.~\eqref{eq:Floquet}, by $P(t)$, i.e., $(\cdot)\mapsto P(t) (\cdot) P^\dagger(t)$, and separating out the lab-frame Hamiltonian, we arrive at Eq.~\eqref{eq:Floquet_STA}
\begin{equation*}
    H(t) = \HF[t] + \AF(t)\,,
\end{equation*}
where $\HF[t]=P(t)\HF P^\dagger(t)$ and $\AF(t)=i \left[\partial_t P(t)\right]P^\dagger(t)$ are periodic since $P(t)$ is periodic.

The key insight to arrive at the Kato decomposition is that $\AF$ is an adiabatic gauge potential for the Floquet states $\ket{\psi_n[t]}$ upon changing the Floquet gauge $[t]$; this is evident from the matrix elements $\bra{\psi_n[t]}\AF(t)\ket{\psi_m[t]}=\braket{\psi_n[t]}{i\partial_t \psi_m[t]}$.
This proves that the lab frame Hamiltonian implements counterdiabatic driving for the Floquet states.
Therefore, by removing the diagonal elements from the Floquet gauge potential, $\AF\mapsto\AK=\AF - \sum_{n} \expval{\AF}_{\psi_n[t]} \ketbra{\psi_n[t]}$, and moving them to the Floquet Hamiltonian, $\HF\mapsto H_\K=\HF + \sum_{n} \expval{\AF}_{\psi_n[t]} \ketbra{\psi_n[t]}$, we obtain the Kato gauge potential $\AK$ and the corresponding Kato Hamiltonian $H_\K$, satisfying $H(t)=H_\K(t)+\A_\K(t)$; this proves the existence of a Kato decomposition.
Moreover, note that the Floquet Hamiltonian is unique up to shifts in the quasienergies by multiples of the driving frequency, $\EF \mapsto \EF+\ell \omega$, $\ell\in\mathbb{Z}$. For Eq.~\eqref{eq:Floquet_STA} to hold in general, these shifts therefore also appear with opposite sign in the diagonal elements of $\AF$; thus, imposing vanishing diagonal elements in $\AK$ removes this gauge-freedom, proving the uniqueness of the Kato decomposition.

\subsection{Proof based on the adiabatic theorem}

Let us now provide a second, independent, proof based on quantum geometry alone, i.e., without using Flqouet's theorem.
To this end, we consider the defining equations for the Kato decomposition
\begin{equation}
    \label{eq:KatoEquations}
    \begin{aligned}
        H(t) &= \HK(t) + \AK(t)\\
        0    &= i\comm{G(\AK(t),\HK(t))}{\HK(t)} \\
        0    &= \bra{\psi_n[t]} \AK(t) \ket{\psi_n[t]}
    \end{aligned}
\end{equation}
where $G(\AK,\HK)=i\comm{\HK}{\AK}-\partial_t \HK$; the second equation is the defining equation of an AGP~\cite{Sels2017_LCD} for $\HK$, and the last equation ensures the parallel-transport properties of $\AK$.

Let us reemphasize that $\HK(t)$ is not a dynamical invariant, since its eigenstates $\EKn{n}(t)$ are explicitly time-dependent.
However, since the $\EKn{n}(t)$ are periodic, we can decompose them, $\EKn{n}(t)=\EKavgn{n}(t) + \EKn{n}^\circ(t)$, into a static part $\EKavgn{n}$ and a periodic part $\EKn{n}^\circ$ with vanishing period average, i.e., $\int_0^T \EKn{n}^\circ(t)\dd t/T=0$.
Following this decomposition of eigenvalues, we can decompose $\HK$ uniquely into two operators $\HKavg[t]$ and $\HK^\circ(t)$ with the same eigenbasis as $\HK$ but eigenvalues $\EKavgn{n}$ and $\EKn{n}^\circ(t)$, i.e.,
\begin{equation}
\label{eq:decomposeKato}
    \HK(t) = \HKavg[t] + \HK^\circ(t)\,,
\end{equation}
with a dynamical invariant $\HKavg[t]$.
Using this decomposition and the defining equations~\eqref{eq:KatoEquations}, we obtain the defining equations for the dynamical invariant $\HKavg[t]$
\begin{equation}
 \label{eq:KatoAverageEquations}
    \begin{aligned}
        H(t)  &=  \HKavg[t] + \AKavg(t) \,,\\
        0 &= i\comm{G(\AKavg,\HKavg)}{\HKavg(t)} \,,\\
        0 &= \int_0^T \bra{\psi_n[t]} \AKavg(t) \ket{\psi_n[t]} \frac{\dd t}{T}\,,
    \end{aligned}
\end{equation}
where $\AKavg=\AK+\HK^\circ(t)$ which justifies the first and third relation.
Moreover, $\AKavg$ is also an adiabatic gauge potential for the states $\ket{\psi_n[t]}$, which are also the eigenstates of $\HKavg$, justifying the second relation.
Note that, $\AKavg$ naturally arises as the Floquet adiabatic gauge potential with respect to the phase of the drive~\cite{Schindler2024_FCD} in the \textit{Floquet-Kato} gauge: This can be best understood in a Samb\'e space/ frequency lattice~\cite{Sambe1973_AtomLightPeriodic} treatment where the periodic time-dependency of Floquet states is  promoted to its own degree of freedom, $\ket{\psi_n[t]}=\sum_{\ell\in\mathbb{Z}} e^{i\ell \omega t} \ket{\psi_{n,\ell}} \mapsto |\psi_n\rangle\rangle = \sum_{\ell\in\mathbb{Z}} \ket{\ell} \otimes \ket{\psi_{n,\ell}}$; then, the inner product for operators $\hat{O}$ on Samb\'e space naturally takes the form  $\langle\langle \psi|\hat{O}|\psi_n\rangle\rangle = \int_0^T \bra{\psi_n[t]} O(t) \ket{\psi_n[t]} \frac{\dd t}{T}$.
Therefore, the last relation in Eq.~\eqref{eq:KatoAverageEquations} corresponds to the usual parallel-transport property, 
$ \langle\langle \psi|\hat{\A}_\K|\psi_n\rangle\rangle $, of the Kato AGP in Samb\'e space.

In the following, we will prove the existence and uniqueness for the dynamical invariant $\HKavg[t]$, as they stand in a one-to-one relation for a given $H(t)$ this also proves the existence and uniqueness of $\HK(t)$:
Note that, above we defined how to obtain $\HKavg[t]$ from $\HK(t)$ uniquely; conversely, $\HK(t)$ can be obtained from $\HKavg[t]$ uniquely by using that they share the same eigenbasis, and $\EKn{n}(t)=\expval{H(t)}_{\psi_n[t]}$.

Using the first identity in Eqs.~\eqref{eq:KatoAverageEquations} to express the $\HKavg[t]$ in terms of $\AKavg[t]$ and the lab-frame Hamiltonian in the definition of $G(\AKavg,\HKavg)$, we can express the first two equations as~\cite{Schindler2024_FCD}
\begin{equation}
    \label{eq:KatoEquationsV2}
    \begin{aligned}
        \HKavg[t] &= H(t) - \AKavg(t) \,,\\
        0      &= \mathcal{L}\pqty{\mathcal{L}\pqty{\AKavg} - \partial_t H}\,,
    \end{aligned}
\end{equation}
where we introduced the linear operator 
$$\mathcal{L}(\cdot)=i\comm{H(t)}{\cdot} + \partial_t\pqty{\cdot}\; ,$$ and used in the second equality that $\HKavg[t]$ is a dynamical invariant, i.e., $\mathcal{L}\pqty{\HKavg}=0$.
Therefore, since the lab frame Hamiltonian $H(t)$ is the known object, we only need to solve for $\AKavg(t)$ to find the Kato decomposition.

We can interpret Eq.~\eqref{eq:KatoEquationsV2} as an implicit equation for $\AKavg(t)$ in terms of the lab-frame Hamiltonian $H(t)$ and its derivative $\partial_t H(t)$.
Indeed, we will now argue that $\A_{+}=\mathcal{L}^+\pqty{\partial_t H}$, where $\mathcal{L}^+$ is the unique Moore-Penrose pseudoinverse of $\mathcal{L}$, provides the sought after unique solution $\A_{+}=\AKavg$.
Since $\mathcal{L}^+$ is the Moore-Penrose pseudoinverse of a hermitian operator $\mathcal{L}$, we have $\mathcal{L}\mathcal{L}\mathcal{L}^+=\mathcal{L}$; therefore, we find that $\A_{+}$ satisfies Eq.~\eqref{eq:KatoEquationsV2}
\begin{align*}
    \mathcal{L}\pqty{ \mathcal{L}\pqty{\A_+} - \partial_t H } &= \mathcal{L}\pqty{ \mathcal{L}\pqty{\mathcal{L}^+\pqty{\partial_t H}} - \partial_t H } 
    \\
    &= \mathcal{L}\pqty{\partial_t H} - \mathcal{L}\pqty{\partial_t H} = 0\, .
\end{align*}
Therefore, $\A_+$ is a valid AGP for $H_+=H(t)-\A_+$. 

It remains to show that $\A_+$'s diagonal elements average to zero; note that this is not strictly required since we could obtain $\AK$ from $\A_+$ uniquely by removing the corresponding diagonal entries.
Notice that, the kernel of $\mathcal{L}$ corresponds to all dynamical invariants, $O(t)=\sum_n O_n \ketbra{\psi_n[t]}$ with constant $O_n$.
Since the Moore-Penrose Pseudo-Inverse $\mathcal{L}^+$ is chosen exactly such that it maps onto the complement of the kernel of $\mathcal{L}$, $\A_+$ has no constant contribution in the diagonal elements.
Therefore, we have proven the existence of an AGP $\A_+=\AKavg$; moreover, due to the uniqueness of the Moore-Penrose pseudoinverse, this is also the unique AGP $\AKavg$.

Note that, $\mathcal{L}$ is an operator on an infinite dimensional Hilbert space, and hence defining the Moore-Penrose pseudoinverse is a priori not well-posed.
However, since all operators are periodic, we can expand them in Fourier harmonics, $O(t)=\sum_n e^{in\omega t}O_n$, such that $\mathcal{L}$ can be represented as a (infinite) matrix, for which we can straightforwardly generalize the concept Moore-Penrose pseudoinverses; for instance, as any non-singular periodic function admits a Fourier decomposition with coefficients that decay at least algebraically, we can consider defining a Moore-Penrose pseudoinverse by truncating to $N_h$ Fourier harmonics, and then take the limit $N_h\to\infty$.
This provides an explicit procedure to approximate the Kato AGP and thus the Kato Hamiltonian with arbitrary precision.
While on physical grounds we expect this procedure to converge, mathematically the algebraic decay of higher harmonic coefficients is not sufficient to rigorously prove this convergence.
For a rigorous proof on the existence of a pseudo-inverse for $\mathcal{L}$ it remains to show that the image of $\mathcal{L}$ is closed~\cite{Pearl1968_PseudoInverse}; however, this directly follows from $\mathcal{L}$ being a finite sum of (tensor products of) operators with closed image.
Notably, this idea may also be extended to quasi-periodically driven systems, by Fourier expanding the operators in two (or more) incommensurate frequencies.

\section{Kato decomposition for degenerate states}
\label{app:degenerate}

In the main text, we focused on non-degenerate Floquet states for notational convenience. However, the results remain valid in the case of degenerate Floquet states.
Notice that, the Floquet quasienergies, and average energies are constant in time such that Floquet states $\ket{\psi_\F[t]}$ which are degenerate at one instance of time are degenerate for all times $t\in[0,T)$; hence, we can consider degenerate subspaces formed by these states.

Therefore, in full analogy to the adiabatic theorem for degenerate subspaces~\cite{Kato1950_adiabatic}, we can simply replace the eigenstate projectors $\Pi_n[\lambda]=\ketbra{\psi_n[\lambda]}$ by eigensubspace projectors $\Pi_{B}[\lambda] = \sum_{\psi_\ell \in B} \ketbra{\psi_\ell[\lambda]}$ projecting onto the subspace of Floquet states with the same quasienergies; here $B$ is an orthonormal basis of the eigensubspace with eigenvalue $\EFn{n}$, i.e., $\EFn{\ell}=\EFn{k}=\EFn{B}$ and $\braket{\psi_\ell[t]}{\psi_k[t]}=\delta_{kl}$ for all $\psi_\ell,\,\psi_k \in B$.
Then, the adiabatic gauge potential still takes the form of Eq.~\eqref{eq:kato_agp} using the eigensubspace projectors~\cite{Kato1950_adiabatic}
\begin{equation}
\label{eq:kato_agp_deg}
    \A_{\mathrm{K},\lambda}=-\frac{1}{2}\sum_B \comm{\Pi_{B}[\lambda]}{i\partial_\lambda \Pi_{B}[\lambda]} \,,
\end{equation}
where $B$ may also correspond to a non-degenerate subspace, containing only a single state.
Note that, $\A_{\mathrm{K},\lambda}$ is fully off-diagonal in the subspaces, i.e., $\bra{\psi_k[\lambda]}\A_{\mathrm{K},\lambda}\ket{\psi_\ell[\lambda]}=0$ for all $\psi_\ell,\,\psi_k \in B$ and all $B$.

Replacing the eigenstate projectors by eigenspace projectors and using the expression~\eqref{eq:kato_agp_deg} for the Kato AGP, Eq.~\eqref{eq:decomposeKato} and the derivation of the Floquet theorem proceeding Eq.~\eqref{eq:decomposeKato} remain valid.
Upon the same replacement also Eqs.~\eqref{eq:Floquet_STA} and \eqref{eq:Kato_evo} remain valid;
the Kato Hamiltonian $\HK(t)$ and average-energy operator $\Havg[t]$ can be expressed in terms of eigensubspace projectors via
\begin{equation}
\label{eq:Kato_deg}
    \begin{aligned}
         \HK(t) 
            &= \sum_B \EKn{B}(t) \Pi_B[t]\,,\\
            &= \sum_B \Pi_B[t] H(t) \Pi_B[t],
    \end{aligned}
\end{equation}
and
\begin{equation}
\label{eq:AE_deg}
    \begin{aligned}
        \Havg[t] 
            &= \sum_B \Eavg_B \Pi_B[t]\\
            &= \sum_B \Pi_B[t] \left[\int_0^T U^\dagger(t^\prime) H(t^\prime) U(t^\prime) \frac{\dd t^\prime}{T}\right] \Pi_B[t], \\
    \end{aligned}
\end{equation}
where the sums run over all eigenspaces.

\subsection{Accidental Degeneracies}

Accidental degeneracies occur due to the fine-tuning of model parameters. They are not protected by symmetries and can be lifted by applying a generic perturbation~\footnote{If a symmetry is present, the perturbation has to respect it.}, resulting in a splitting that is linear in the perturbation strength. 
In particular, accidental degeneracies can occur in the quasienergy spectrum $\EF$  due to folding, for fine-tuned values of the drive frequency. If this is the case, the Lanczos algorithm for exact diagonalization selects an arbitrary basis within the accidental-degenerate subspace.  

In the case of accidental quasienergy degeneracies,
the average-energies need not be degenerate; in other words, since $\EF=\Eavg+T^{-1}\gamma$, the average-energies can be non-degenerate $\Eavg_{n}\neq\Eavg_{m}$ but the sum of average-energy and geometric phase can still be degenerate: $\Eavg_{n}+T^{-1}\gamma_n =\Eavg_{m}+T^{-1}\gamma_m$.
In such cases, further diagonalizing of the average-energy operator~\eqref{eq:AE_deg} lifts the degeneracy.
These accidental degeneracies can occur in many-body systems due to quasi-energy folding; for instance, they occur throughout the non-heating regime of the mixed-field Ising model, Fig.~\ref{fig:floquetGS}, and near the transition in the discrete time crystal, $(\pi-\theta_x)\approx 0.02\pi$.

\subsection{Degeneracies due to Symmetries}

If $H(t)$ admits a unitary symmetry, i.e., $\comm{\mathcal{S}}{H(t)}=0$ for all $t\in[0,T)$ with symmetry operator $\mathcal{S}\mathcal{S}^\dagger=\identity$, degeneracies due to this symmetry in $\UF$ also cause degeneracies in the average energy. 
Specifically, if $\ket{\psi_{\F,n}[t]}$ and $\ket{\psi_{\F,m}[t]}=\mathcal{S}\ket{\psi_{\F,n}[t]}$ are two distinct eigenstates of $\UF$ connected by the symmetry, they share not only the same quasi-energy, but also the same average-energy, since
\begin{align*}
    \Eavg_n   
        &= \int_0^T \bra{\psi_{\F,n}[t]} H(t)\ket{\psi_{\F,n}[t]} \frac{\dd t}{T} \\
        &= \int_0^T \bra{\psi_{\F,n}[t]}\mathcal{S}^\dagger H(t) \mathcal{S}\ket{\psi_{\F,n}[t]} \frac{\dd t}{T}\\
        &= \int_0^T \bra{\psi_{\F,m}[t]} H(t)\ket{\psi_{\F,m}[t]} \frac{\dd t}{T} \\
        &= \Eavg_m\,.
\end{align*}

However, the converse is not true: a degenerate average-energy spectrum does not imply degeneracies in the quasienergy. An example is the discrete time crystal model discussed in the main text, where the quasienergy spectrum is non-degenerate due to the $\theta_x$-kick. In turn, this also implies nondegenerate geometric phases, due to $\EF=\Eavg+T^{-1}\gamma$.

Notably, spontaneous symmetry breaking (SSB) can occur in the average-energy spectrum in much the same way as it does in Hamiltonian spectra in equilibrium.
In the simplest scenario, this requires a doublet of eigenstates, containing the ground and first excited states that transform into each other under the symmetry, with an average-energy gap that closes in the thermodynamic limit.
However, unlike equilibrium systems where SSB is tied to spatial symmetries, in Floquet systems SSB can also occur in time: e.g., in time-crystalline phases, the gap closing in the average-energy is tied to a $\pi$-gap `opening' in the geometric phases.

Finally, let us emphasize that any symmetry of $H(t)$, $\comm{\mathcal{S}}{H(t)}=0$ $\forall t{\in}[0,T)$, is also inherited by the Floquet unitary, $[\UF,\mathcal{S}]=0$.
Therefore, it is often more convenient to restrict numerical and analytical computations to symmetric subspaces; this approach lifts the degeneracies related to the symmetries and reduces the Hilbert space dimension.
For instance, we followed this strategy for the mixed-field Ising and discrete time crystal example, where we restrict to fixed momentum blocks and parity sectors, respectively.

\section{Proof of the relations in Table~\ref{tab:gauges}}
\label{app:proof_table}

Here, we derive the accumulated phases for the different counterdiabatic gauges in Table~\ref{tab:gauges}; for the phase accumulated during adiabatic evolution see, e.g., Refs.~\cite{Born1928_adiabatic,Kato1950_adiabatic}.

Let us recap the derivation of CD driving: For a control Hamiltonian $\Hctrl(\lambda)$ with transformation $V_\lambda$ to the instantaneous eigenbasis, the CD Hamiltonian is $\HCD=\Hctrl+\dot{\lambda}\A_\lambda$ with the \textit{dynamical} AGP $\A_{\mathrm{D},\lambda}=\left(i\partial_\lambda V_\lambda\right)V^\dagger\lambda$.
Therefore, in the instantaneous eigenframe of $V_\lambda$, the dynamics generated by $\Hctrl+\A_{\mathrm{D},\lambda}$ are described by $-i \partial_t \ket{\Tilde{\psi}(t)} = \Tilde{H}_\ctrl \ket{\Tilde{\psi}(t)}$ with $\Tilde{H}_\ctrl$ diagonal; hence, the time-evolution starting from an eigenstate of $\Tilde{H}_\ctrl$ is simply given by $\ket{\Tilde{\psi}(t)} = e^{i\phi_n(t)} \ket{\Tilde{\psi}_n}$.
Going back to the lab frame, the full evolution starting from an eigenstate of $\Hctrl$, is thus described by
\begin{equation*}
    \ket{\psi(t)} = V_\lambda  e^{i\phi_n(t)} \ket{\Tilde{\psi}_n} = e^{i\phi_n(t)} \ket{\psi[\lambda(t)]}\, ,
\end{equation*}
proving the phase accumulation for the dynamical CD driving.

To derive the accumulated phases in the other rows of Table~\ref{tab:gauges}, let us recall that different gauge choices only differ in diagonal elements, i.e., $\A^\prime_\lambda=\A_\lambda + \mathcal{D}(\lambda)$, where $\comm{\mathcal{D}(\lambda)}{\Hctrl(\lambda)}=0$ and $\mathcal{D}=\sum_n \partial_\lambda \chi_n(\lambda) \ketbra{\psi_n[\lambda]}$.
Therefore, $V_\lambda$ also transforms to the instantaneous eigenbasis of $\Hctrl-\mathcal{D}$; thus, by decomposing $\Hctrl+\A^\prime_\lambda=\pqty{\Hctrl-\mathcal{D}}+\A_{\mathrm{D},\lambda}=\Hctrl^\prime + \A_{\mathrm{D},\lambda}$, with the dynamical AGP $\A_{\mathrm{D},\lambda}=\left(i\partial_\lambda V_\lambda\right)V^\dagger\lambda$, the results in Table~\ref{tab:gauges} follow immediately from retracing the steps from above, by noting that $\phi_n(t)$ needs to be replaced by $\phi_n(t)+\chi_n(t)$.

For the Kato formulation in the parallel-transport gauge, specifically, we note that $\partial_\lambda \chi_n = -\braket{\psi_n}{i\partial_\lambda \psi_n}$, and thus $\phi_n(t)\mapsto \phi_n+\gamma_n(t)$, where we identified the geometric phase $\gamma_n(t)=-\int_{\lambda(0)}^{\lambda(t)}\braket{\psi_n[\lambda]}{i\partial_\lambda \psi_n[\lambda]}\dd \lambda$.

\section{Infinite frequency limit of the Average-Energy Operator and High-frequency expansion for the Kato decomposition}
\label{app:HFE}

Let us consider the infinite frequency limit for the average-energy operator $\Havg$; we will show that $\Havg\to\HF^{(0)}$ as $\omega\to\infty$ where $\HF^{(0)}$ is the average Hamiltonian $\HF^{(0)}=\int_0^T H(t) \mathrm{d}t/T=\overline{H}$, or the zeroth order in the high-frequency expansion.
To this end, it suffices to show that the Berry phase vanishes in the infinite frequency limit, $\gamma_n^{(0)}(T)/T=\int_0^T \bra{\psi_n^{(0)}}\AF^{(0)}\ket{\psi_n^{(0)}}\dd t/T=0+O(T)$.
Intuitively, this seems clear since $\HF^{(0)}$ has no Floquet-gauge dependence; however, one has to be careful with the power counting since the Floquet gauge potential $\bra{\psi_n}\AF(t)\ket{\psi_n}=\braket{\psi_n}{i\partial_t \psi_m}$ contains a derivative $O(T)$, i.e., $\AF^{(0)}\neq0$.
In fact, since $H(t)$ is of order $O(T^0)$, the identity $H(t)=\HK(t)+\AK(t)=\HF(t)+\AF(t)$ also holds in the high-frequency expansion~(HFE) to leading order, i.e., for HFE up to order $T^n$ this relation is satisfied up to errors $O(T^{n+1})$.
Therefore, we have $\AF^{(0)}(t)=H(t)-\overline{H}$; this is in agreement with the observation that the Floquet eigenstates change in time in the next order $O(T)$, i.e., $\ket{\psi_n[t]}=\ket{\psi_n^{(0)}}+O(T)$. Therefore, we find
\begin{align*}
    T^{-1}\gamma_n  &= \int_0^T \bra{\psi_n^{(0)}} H(t)-\overline{H} \ket{\psi_n^{(0)}}\dd t/T + O(T)\\
                    &= \bra{\psi_n^{(0)}}  \int_0^T  H(t)-\overline{H} \dd t/T \ket{\psi_n^{(0)}}+ O(T)\\
                    &= \bra{\psi_n^{(0)}}  \left[ \overline{H}-\overline{H} \right]\ket{\psi_n^{(0)}}+ O(T)\\
                    &= 0+ O(T) \, ,
\end{align*}
and hence, $\Eavg_n=\EFn{n}+O(T)$, and the Berry phase is generally of order $O(T)$ as $T\to0$ (which points at its inherently nonequilibrium character).
However, for higher-order terms in the period $T$, the Berry phase is non-negligible as is also evident from the generic Floquet-gauge dependence of the high-frequency Hamiltonian.

For a general high-frequency expansion to order $O(T^n)$ of the average-energy operators, one can consider the following procedure: 
\begin{enumerate}
    \item compute the HFE Floquet Hamiltonian $\HF^{(n)}[t]$, and $\AF^{(n)}(t)=H(t)-\HF^{(n)}[t]$;
    \item compute the eigenstates $\ket{\psi_m^{(n)}[t]}$ of $\HF^{(n)}[t]$;
    \item finally, the Kato energies are obtained from $\expval{\HK^{(n)}(t)}_{\psi_m^{(n)}[t]}=\expval{H(t)}_{\psi_m^{(n)}[t]}$, or via the Kato AGP $\AK(t)=-\frac{1}{2}\sum_m \comm{\Pi_m^{(n)}[t]}{i\partial_t \Pi_m^{(n)}[t]}$ with $\Pi_m^{(n)}[t]=\ketbra{\psi_m^{(n)}[t]}$
\end{enumerate}
It is not evident that the crucial step 2 can be cut short using a high-frequency expansion; also we are unaware of any high-frequency expansion for the Berry phase or the Wilson-line operator.

\begin{figure}[t]
    \centering
    \includegraphics[width=\linewidth]{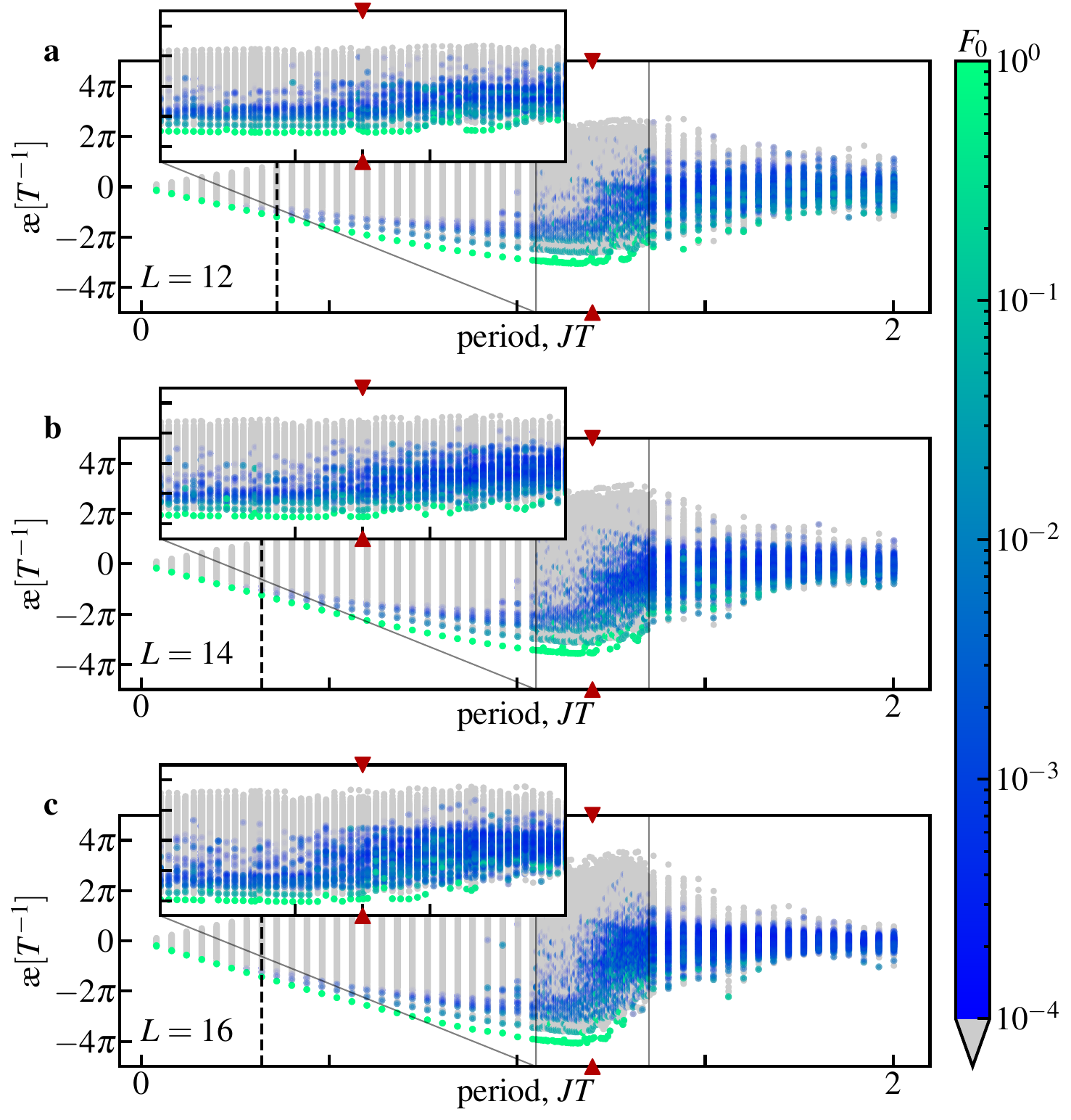}
    \caption{
        \textbf{System size scaling of Floquet ground state overlap in kicked mixed-field Ising chain}.
        \textbf{a}-\textbf{c}, average-energy spectrum for $L=12,14,16$ as function of period, respectively.
        \textit{Inset} shows zoom into $JT\in[1.15,1.25]$ region.
        Colorbar indicates overlap $F_0=\abs{\braket{\psi_n}{\psi_0}}^2$ with ground state $\ket{\psi_0}$ of $\HF^{(0)}$.
        There seems to be a slight shift in the location of the resonance towards smaller periods with increasing system size. Notably, the spectral bandwidth increases with system size before the transition point, but decreases after the transition point.
    }
    \label{fig:MFI_L}
\end{figure}

\begin{figure}[t]
    \centering
    \includegraphics[width=\linewidth]{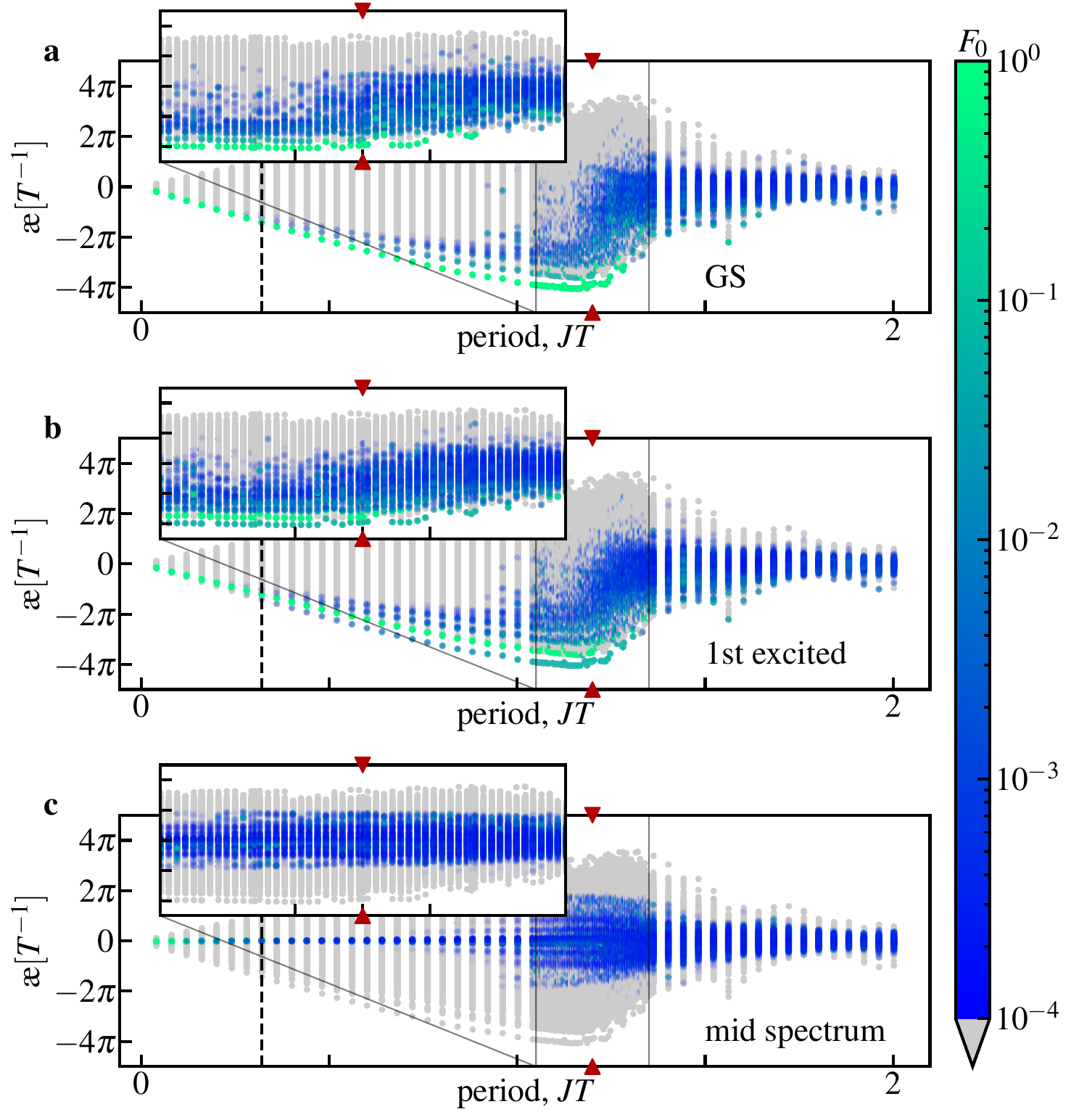}
    \caption{
        \textbf{State dependence of overlap in kicked mixed-field Ising model}.
        \textbf{a}-\textbf{c}, average-energy spectrum for the ground state as a function of the period, with the colorbar indicating the overlap with the ground state, first excited state and mid-spectrum eigenstate of $\HF^{(0)}$, respectively.
        \textit{Inset} shows zoom into $JT\in[1.15,1.25]$ region.
        The resonance proliferation point, where the perturbative Floquet states start deviating strongly from the exact Floquet states, appears at smaller periods when moving away from the ground state; this indicates that the ground state resists Floquet heating for larger periods compared to the excited Floquet states.
    }
    \label{fig:MFI_state}
\end{figure}


\section{Further details on XY model example}
\label{app:xy}

In this appendix, we explicitly derive the free fermion model~\eqref{eq:XY_fermion} from the spin-$\frac{1}{2}$ XY-model~\eqref{eq:XY}. We also derive the Floquet Hamiltonian for the resulting ensemble of two-level systems.

Recall the XY-model Hamiltonian, Eq.~\eqref{eq:XY},
\begin{equation*}
    H(t) = \frac{1}{2}\sum_{n=1}^L \left[ \left( J \sigma^+_{n+1} \sigma^-_{n} + A i e^{-i\omega t} \sigma^+_{n+1} \sigma^+_{n} + \mathrm{h.c.}\right) + \frac{g}{2} \sigma_n^z\right]\,,
\end{equation*}
with periodic boundary conditions, $L{+}1{\equiv}1$, spin-lowering and raising operators $\sigma^\pm_\ell=\pqty{\sigma^x_\ell \pm  i \sigma^y_\ell}/2$ and pauli-matrices $\sigma^\alpha$, $\alpha=x,y,z$.
This model is exactly solvable by mapping it to a free-fermion model via a Jordan-Wigner transformation~\cite{Barouch_etal_XY_1971,Fradkin1978_XY}
\begin{equation}
    \begin{aligned}
        f_\ell         &= \prod_{j=1}^{\ell-1}\pqty{-\sigma^z_j} \sigma_\ell^{-}\\
        f_\ell^\dagger &= \prod_{j=1}^{\ell-1}\pqty{-\sigma^z_j} \sigma_\ell^{+}\\
        f^\dagger_\ell f_\ell &= \frac{1}{2}\pqty{\sigma^z_\ell +1} \,,
    \end{aligned}
\end{equation}
which results in the real-space free fermion model
\begin{equation*}
    H(t) \hat{=} \frac{1}{2}\sum_{n=1}^L \left[ \left( J f^\dagger_{n+1} f_{n} + A i e^{-i\omega t} f^\dagger_{n+1} f^\dagger_{n} + \mathrm{h.c.}\right) + g f^\dagger_n f_n\right]\,,
\end{equation*}
up to an irrelevant constant additive factor.
Performing, additionally, a Fourier transform $f_\ell^{\dagger}=L^{-1/2}\sum_k e^{ik\ell}f_k^\dagger$ we arrive at Eq.~\eqref{eq:XY_fermion},
\begin{equation*}
    \begin{aligned}
        H(t)    & \hat{=} \sum_k \boldsymbol{\psi}_k^\dagger h(k,t) \boldsymbol{\psi}_k\, ,\\
        h(k,t)  &=
            \Delta_k\tau^z + A_k\left[ \cos(\omega t) \tau^x + \sin(\omega t) \tau^y \right]\, ,
    \end{aligned}
\end{equation*}
where the momentum-dependent level splitting is $\Delta_k=g + J \cos(k) $, drive amplitude is $A_k=A\sin(k)$, the spinor is $\boldsymbol{\psi}_k^\dagger=\left( f_k^\dagger,\, f_{-k} \right)$, and the pseudo-spin operators are $\tau^\alpha=\sigma^\alpha/2$. 

Since the quasi-momentum modes decouple, the system reduces to a collection of two-level systems described by the Bloch Hamiltonian $h(k,t)$ at fixed momentum $k$.
The time dependence can, thus, be removed by transforming to a rotating frame $V_1(t)=\exp[-i\omega ( t - t_0)\tau^z]$ for each $k$, resulting in
\begin{equation}
    \label{eq:XY_rot}
    h^\mathrm{rot}_k[t_0] = \left[\Delta_k-\omega\right]\tau^z + A_k \left[ \cos(\omega t_0) \tau^x + \sin(\omega t_0) \tau^y \right]\, .
\end{equation}
Note that, the eigenvalues $\epsilon^\mathrm{rot}_{k;1,2}=\pm \epsilon_k/2$, $\epsilon_k^2=\left[\Delta_k-\omega\right]^2+A_k^2$ of $h^\mathrm{rot}_k$ are independent of the initial time $t_0$, and diverge in the infinite frequency limit, $\epsilon^\mathrm{rot}_{k;1,2}\overset{\omega\to\infty}{\longrightarrow}\pm \omega$.
Thus, to obtain a Floquet Hamiltonian that reproduces the correct infinite frequency limit, we have to shift the energies by performing a second rotating frame transformation, with $V_2(k,t_0)=\exp[-i\omega t_0 h^\mathrm{rot}(k)/\epsilon]$, to obtain
\begin{equation*}
    h_{\F,k}[t] = \pqty{\epsilon_k - \omega} \left\{ \frac{\Delta_k-\omega}{\epsilon_k} \tau^z + \frac{A_k}{\epsilon_k} \left[ \cos(\omega t) \tau^x + \sin(\omega t) \tau^y \right]  \right\} \, ,
\end{equation*}
where we also replaced $t_0$ by $t$.
Using the identity $H(t)=H_\mathrm{F}(t)+\AF(t)$, or the fact that $P(t)=V_1(t)V_2(t)$, we can directly obtain
\begin{equation*}
    a_{\F,k}(t) = \omega\Bqty{ \pqty{1 + \frac{\Delta_k-\omega}{\epsilon_k}} \tau^z + \frac{A_k}{\epsilon_k} \left[ \cos(\omega t) \tau^x + \sin(\omega t) \tau^y \right]}\, .
\end{equation*}
Using the eigenstate projectors, $\Pi_{\pm}=\frac{1}{2}\pqty{\identity\pm h_{\F,k}[t]}/(\epsilon_k-\omega)$, and $a_{\F,k}(t)=\frac{1}{2}\sum_n\comm{\Pi_n}{i\partial_t \Pi_n}$, we can also readily compute the Kato Hamiltonian and gauge potential
\begin{equation}
    \begin{aligned}
        h_{\K,k}(t) &= \frac{\epsilon_{\K,k}}{\epsilon_k} \left\{ \frac{\Delta_k-\omega}{\epsilon_k} \tau^z + \frac{A_k}{\epsilon_k} \left[ \cos(\omega t) \tau^x + \sin(\omega t) \tau^y \right]  \right\} \, ,\\
        a_{\K,k}(t) &= \frac{A_k\omega}{\epsilon_k} \left\{ \frac{A_k}{\epsilon_k} \tau^z -  \frac{\Delta_k-\omega}{\epsilon_k}\left[ \cos(\omega t) \tau^x + \sin(\omega t) \tau^y \right]  \right\} \, ,
    \end{aligned}
\end{equation}
with $\epsilon_{\K,k}=\left(\Delta_k-\omega\right)\Delta_k+A_k^2$. 

Taking a closer look at the resonance condition for the Floquet energies, $\Delta_k=\omega$, and the position of the average-energy crossing, $\epsilon_{\K,k}=0$, we find that the exact level crossing of the average-energies appears strictly before the Floquet resonance. Thus, the level-crossing is indicative of a photon resonance but does not happen exactly on resonance.
A precise interpretation of this crossing point is an interesting open question for future studies.

\begin{figure}[t]
    \centering
    \includegraphics[width=\linewidth]{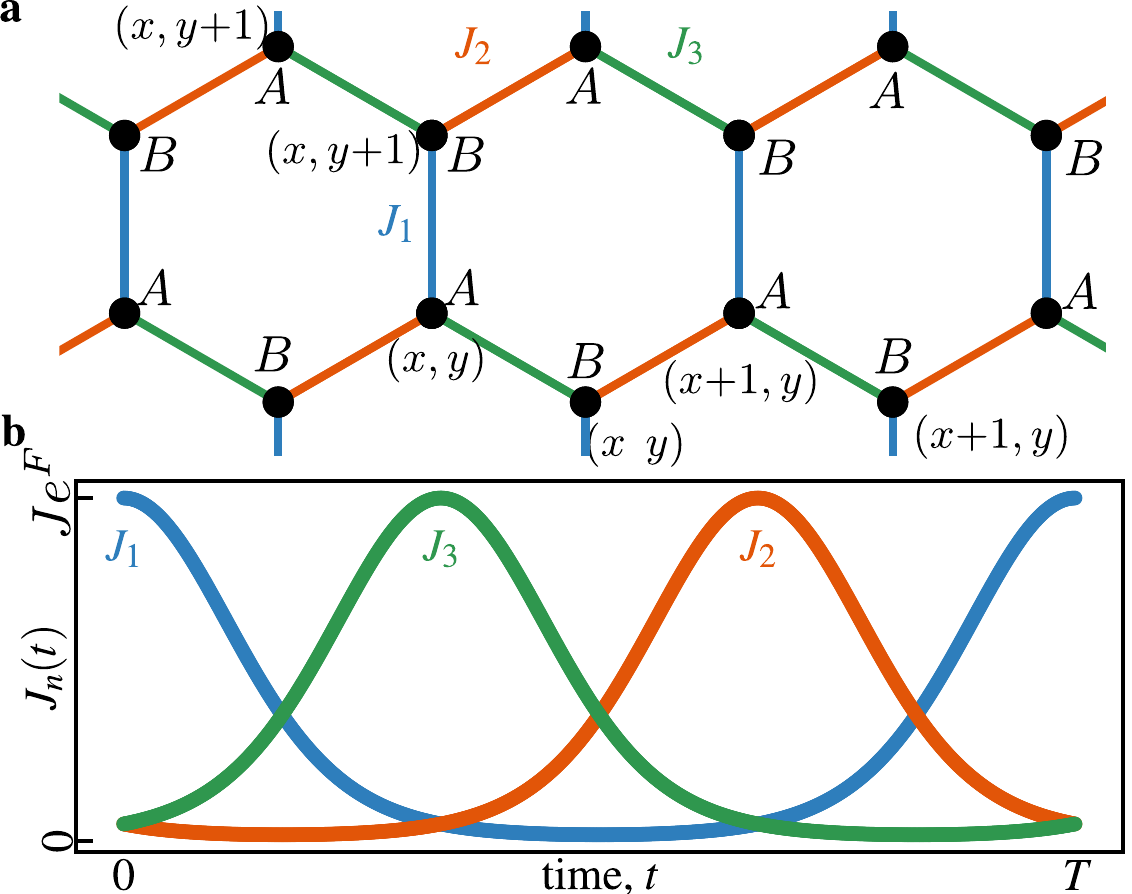}
    \caption{
        \textbf{Anomalous Floquet Topological Insulator model}, extended version of Fig.~\ref{fig:AFTI}a.
        \textbf{a}, Sketch of hexagonal lattice, with $A$, $B$ sub-lattice and $(x,y)$ coordinates indicated.
        \textbf{b}, time-dependence of hopping amplitudes $J_{1,2,3}(t)$.
    }
    \label{fig:AFTI_model}
\end{figure}

\section{Additional information for kicked mixed-field Ising Floquet ground state example}
\label{app:floquetgs}

\subsection{Definition of locality measure}

In Fig.~\ref{fig:floquetGS}c and d, we compare the locality of the Floquet Hamiltonian $\HF$ and average-energy operator $\Havg$.
As a measure of locality we consider the relative weight of one- and two-body operators, which we define as follows:
\begin{equation}
    \mathrm{locality}(O) = \frac{\norm{O_{\text{1-body}} + O_{\text{2-body}}}_2}{\norm{O}_2} \, ,
\end{equation}
where $O_{\text{1-body}}$ and $O_{\text{2-body}}$ are the one- and two-body terms entering the operator $O$, respectively, and $\norm{O}_2 = \sqrt{\Tr(O^2)}$ is the Frobenius norm.
We choose to include only up to two-body operators for easier computability; however, we checked that including higher body terms, e.g., including three or four body operators, does not qualitatively change the results.

More formally, we can decompose any (tracelesss) operator $O$ into Pauli-strings
\begin{equation}
    O = \sum_{n=1}^L \sum_{\alpha_n\in\Bqty{0,x,y,z}} c_{\boldsymbol{\alpha}} P_{\boldsymbol{\alpha}}
\end{equation}
with concatenated index $\boldsymbol{\alpha}=(\alpha_1,\dots,\alpha_L)$, coefficients $c_{\boldsymbol{\alpha}}$, and Pauli operators,
\begin{equation}
    P_{\boldsymbol{\alpha}} = \bigotimes_{n=1}^L \sigma_{n}^{\alpha_n}
    \label{eq:pauli_string}
\end{equation}
where $\alpha_n=0,x,y,z$, and $\sigma_n^0=\identity$ is the identity.
Then, $\mathcal{P}_\ell$ is the set of $\ell$-body Pauli operators, i.e., all Pauli-strings~\eqref{eq:pauli_string} where exactly $\ell$-many $\alpha_n$ are not equal to $0$; likewise, the operator $O_{\ell\text{-body}}$ corresponds to the operator $O$ restricted to $\ell$-body Pauli-strings
\begin{equation}
    O_{\ell\text{-body}} = \sum_{P \in \mathcal{P}_\ell} \frac{\Tr(PO)}{\norm{P}_2^2} P\, .
\end{equation}
Note that, without restricting to symmetries the $\mathcal{P}_\ell$ has on the order $O\pqty{3^\ell \binom{L}{\ell}}$ many operators, severely restricting the accessible system sizes.

\subsection{Additional Data}
In Fig.~\ref{fig:MFI_L}, we show the system size dependence of the putative heating phase transition in the kicked mixed-field Ising model, cf.~discussion of Fig.~\ref{fig:floquetGS}.
We observe that the first photon resonance encountered in the system shifts to smaller periods, and the number of photon resonances encountered increases with increasing system size.
More notably, the many-body bandwidth of the average-energy spectrum appears to increase before the transition point and decrease after the transition point with increasing system size; this is indicative of an extensive energy spectrum before the transition point which suggests the existence of a local Hamiltonian description.
Whether the putative heating transition is related to a locality-to-nonlocality transition in the average-energy operator is an interesting question for future studies.

Moreover, in Fig.~\ref{fig:MFI_state}, we study the overlap of different states of the infinite frequency, average Hamiltonian, with the exact Floquet states as we vary the period.
In agreement with earlier studies~\cite{Ikeda_2024}, we find that when moving away from the ground state the perturbative eigenstates gain overlap with a large fraction of the exact Floquet eigenstates already at lower period values.

Let us emphasize that to make clear-cut statements about the existence of a heating phase transition, larger system sizes and a more rigorous finite size scaling are necessary, which are beyond the scope of this work.
However, the Kato decomposition opens up new avenues that could potentially detect heating transitions, e.g., from the extensivity of the spectrum or the locality of the average-energy operator.

\begin{figure}[t]
    \centering
    \includegraphics[width=\linewidth]{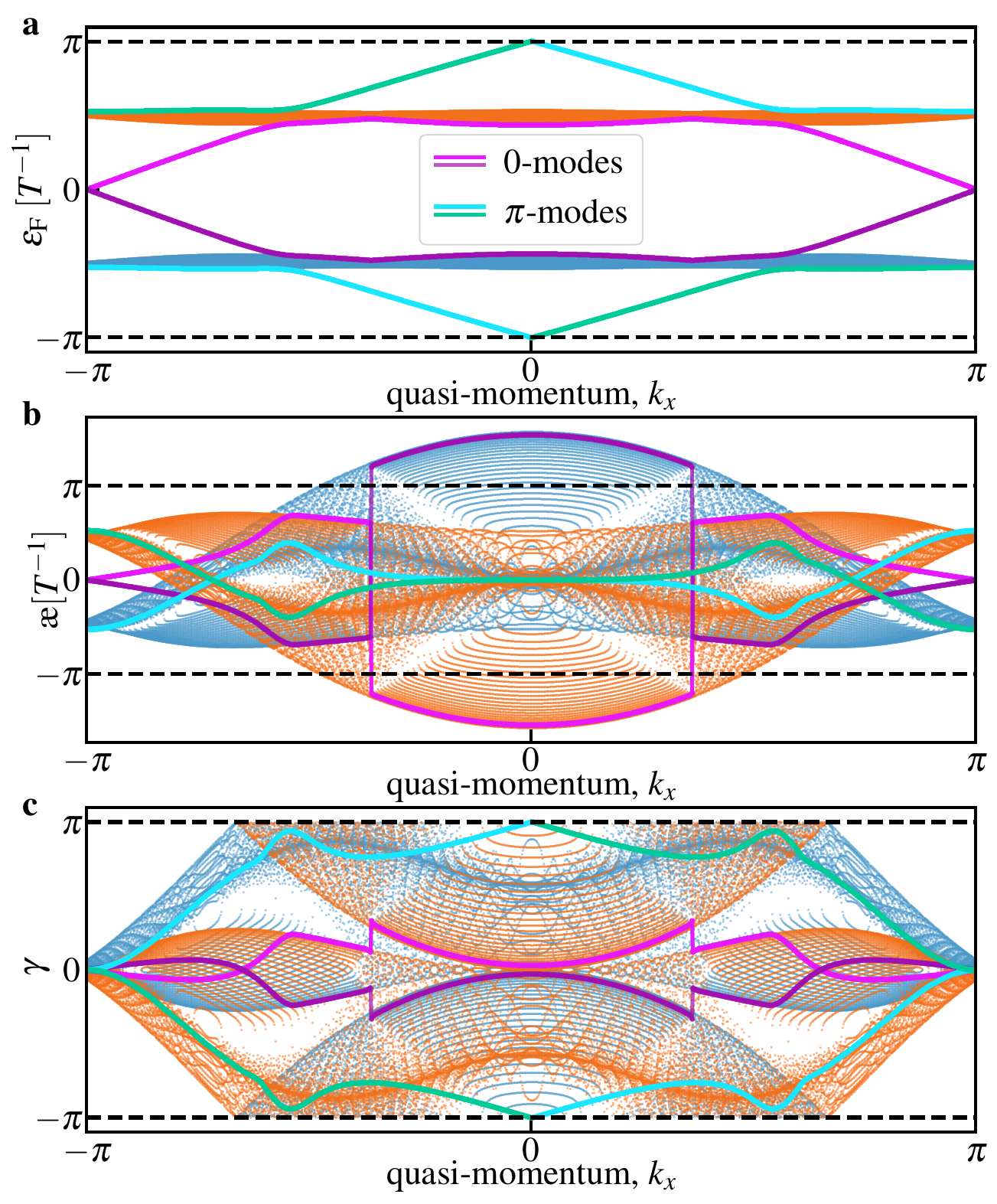}
    \caption{
        \textbf{Anomalous Floquet Topological Insulator, extended data.}
        \textbf{a}-\textbf{c}, Floquet, average-energy, and closed Wilson line spectrum as functions of the quasi-momentum $k_x$, respectively; same data as in Fig.~\ref{fig:AFTI}.
        Magenta and cyan lines indicate the two $0$ edge modes and two $\pi$ edge modes, respectively.
        The blue (orange) color corresponds to the (arbitrary) coloring of lower (upper) Floquet band as shown in a.
        In the Kato decomposition, in both the average-energy operator and the Berry phases, a chaotic and regular regime becomes apparent for $\abs{k_x} \lessapprox 4\pi/5$. This sharply contrasts with the regular flat Floquet bands.
    }
    \label{fig:AFTI_detail}
\end{figure}

\begin{figure}[t]
    \centering
    \includegraphics[width=.9\linewidth]{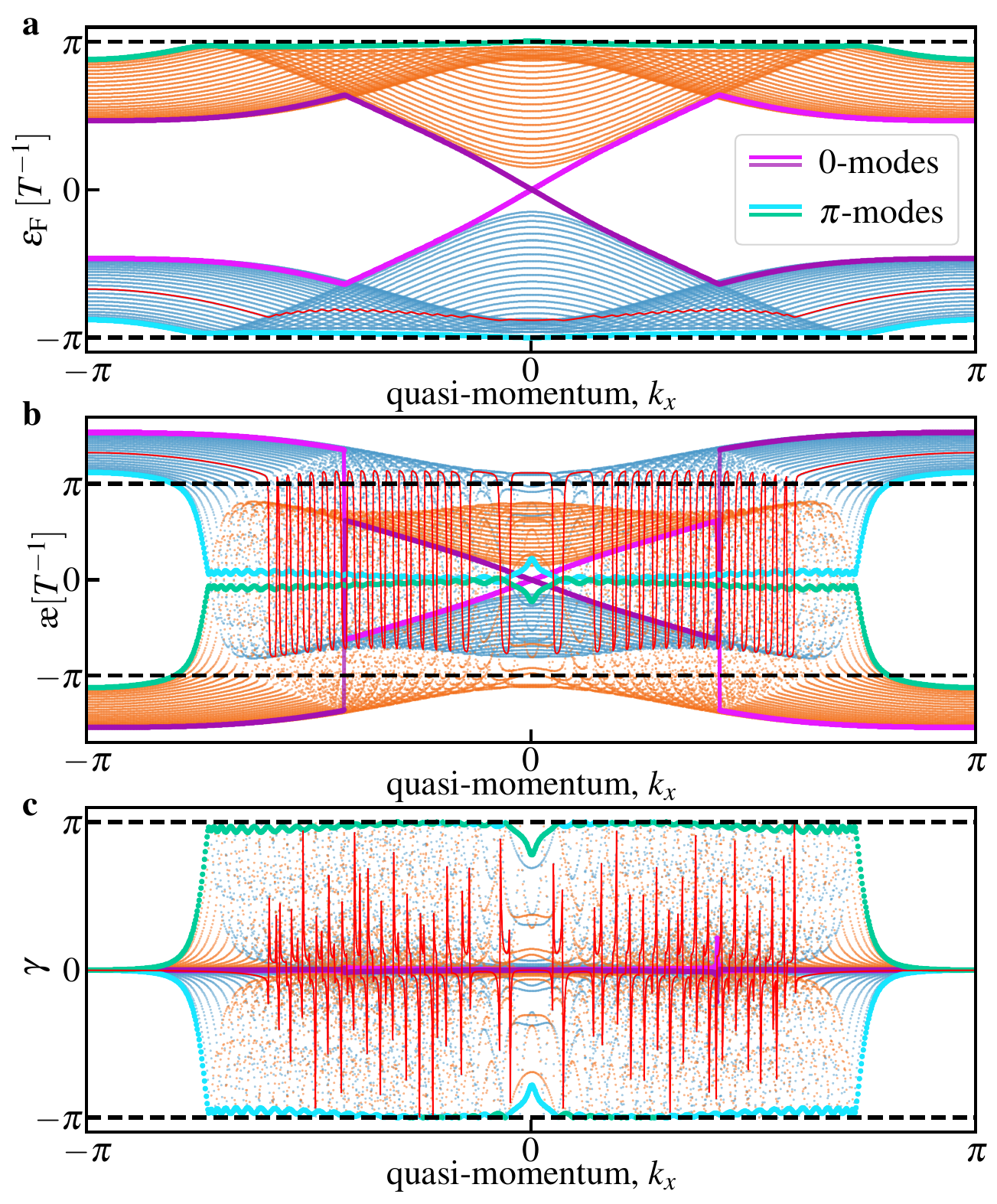}
    \caption{
        \textbf{Anomalous Floquet Topological Insulator, second example}
        \textbf{a}-\textbf{c}, Floquet, average-energy and (folded) Berry spectrum for the model~\eqref{eq:AFTI_Rudner} as functions of the quasi-momentum $k_x$, respectively.
        The blue (orange) color corresponds to the arbitrary coloring of the lower (upper) Floquet band as shown in a.
        Magenta and cyan lines indicate the two $0$ edge modes and two $\pi$ edge modes, respectively; the thin solid red line indicates a representative line from the bulk bands.
        Similar, to the example in the main text, the $0$ and $\pi$ are not distinguishable in the average-energy spectrum.
        However, the bulk bands undergo a multitude of one-photon resonances, resulting in a more chaotic spectrum than for the example in the main text.
        Data shown for $L_x=500$ values, however, for sorting the states $5000$ $k_x$ values have been considered; other parameters are detailed in text following Eq.~\eqref{eq:AFTI_Rudner}
    }
    \label{fig:AFTI_Rudner}
\end{figure}

\section{Anomalous Floquet Topological Insulator examples}
\label{app:AFTI}

In this appendix, we give more details about the anomalous Floquet topological insulator example considered in the main text.

We consider a model of non-interacting fermions on a hexagonal lattice with hopping amplitudes that are varied homogeneously in space but in a chiral, time-periodic way.
Concretely we consider the Hamiltonian~\cite{Dutta2024_AFTI}, cf.~Fig.~\ref{fig:AFTI_model},
\begin{equation}
\label{eq:AFTI_hexagon}
    \begin{aligned}
        H(t) = \sum_{x=1}^{L_x} \sum_{y=1}^{L_y} & \left[ J_1(t) c^\dagger_{x,y,B} c_{x,y-1,A} + J_2(t) c^\dagger_{x,y,B}c_{x+1,y,A}\right.\\
                                                 &\left.  + J_3(t) c^\dagger_{x,y,B}c_{x,y,A} \right] + \mathrm{h.~c.}
    \end{aligned}
\end{equation}
with $c^{(\dagger)}_{x,y,Z}$ annihilating (creating) a fermion at lattice site labeled by $(x,y)$ on the sub-lattice $Z$, $Z=A,B$, with real hopping amplitudes $J_n(t)=J \exp\bqty{F \cos(\omega t + \phi_n)}$ and $\phi_n=(n-1)2\pi/3$; we set the lattice constant to $a=1$, and use $L_x=500$, $L_y=30$, $F=2J$ and $\omega=8.7J$ throughout, unless states otherwise.
We consider cylindrical boundary conditions, with the open boundary in the $y$-direction and the periodic boundary in the $x$-direction; therefore, only the $k_x$ quasi-momentum is a good quantum number.
In Fig.~\ref{fig:AFTI_detail}, we also provide a more detailed view on the data from Fig.~\ref{fig:AFTI}, providing a clearer picture of the chaotic regime appearing both in the average-energy operator and the Berry phase.

To emphasize that our results are not model specific, let us also consider a different AFTI model, studied in Ref.~\cite{Rudner2013_AnomalousFloquetEdgeState}.
This model is described by a general Bloch-Hamiltonian on a bipartite lattice
\begin{equation}
\label{eq:AFTI_Rudner}
    h(\boldsymbol{k},t) = \boldsymbol{d}(\boldsymbol{k},t) \cdot \boldsymbol{\sigma} \,,
\end{equation}
with $d_x(\boldsymbol{k}) = a \sin(k_x)$, $d_y(\boldsymbol{k})=a \sin(k_y)$, and $d_z(\boldsymbol{k},t)=\Delta(t) + (\mu-J) - 2b \left[ 2- \cos(k_x)-\cos(k_y) \right]+J\cos(k_x)\cos(k_y)$, where $\Delta(t)=\Delta_0 \cos(\omega t)$; we consider $J=b=1.5\mu$, $a=4\mu$ and $\Delta_0=\mu$, $\Delta_0/\omega=0.07$, and $L_y=50$.
This model leads to qualitatively similar behavior as the AFTI on the hexagonal lattice~\eqref{eq:AFTI_hexagon}, see Fig.~\ref{fig:AFTI_Rudner}:
the $0$ and $\pi$ edge modes can no longer be distinguished in the average-energy spectrum but require the additional information contained in the Berry phase.
Moreover, here, the average-energy bulk bands undergo multiple one-photon-resonances when varying quasi-momentum $k_x$, leading to `chaotic' behavior of the bulk bands.

\bibliography{FSTA_biblio}
\end{document}